%% file: main.tex
\documentclass[a4paper, 10pt, twocolumn]{article}
 
\input{preamble}

\title{Which wind and solar modelling choices matter? Effects on capacity factors, energy droughts and European power system configuration}
 
\author[1]{L.~Karkossa\thanks{Corresponding author, \url{lalka@dtu.dk}}}
\author[1]{A.~Grochowicz}
\author[1]{M.~J.~Koivisto}
\author[1]{A.~Hahmann}
\author[1]{M.~Victoria}

\affil[1]{Department of Wind and Energy Systems, Technical University of Denmark, 2800 Kgs. Lyngby, Denmark}

\date{\today}
 
\begin{document}
\input{chapters/00_graphical_abstract}
\twocolumn[
\maketitle
\input{chapters/00_abstract}
\vspace{2ex}
]
% \thanks text is lost inside \twocolumn[...]; set it again (keep in sync with \author above)
{\renewcommand{\thefootnote}{\fnsymbol{footnote}}\footnotetext[1]{\hypertarget{Hfootnote.1}{}Corresponding author, \url{lalka@dtu.dk}}}
\input{chapters/01_introduction}

\input{chapters/02_methods}
\input{chapters/03_results}

\FloatBarrier % all Results figures before the Discussion
\input{chapters/04_discussion_conclusion}

\section*{Code and data availability}
All code is open source (licensed under MIT), and all data used are open (various licences). The code and instructions to reproduce the study can be found in the following GitHub repository: \url{https://github.com/LTBK23/vre_modelling_choices}.

\section*{Acknowledgements}
ERA5 reanalysis data~\cite{c3s_era5_2018} were downloaded from the Copernicus Climate Change Service (C3S). The results contain modified Copernicus Climate Change Service information 2025. Neither the European Commission nor ECMWF is responsible for any use that may be made of the Copernicus information or data it contains. The authors gratefully acknowledge the computational and data resources provided on the Sophia HPC Cluster at the Technical University of Denmark~\cite{technical_university_of_denmark_sophia_2019}.

L.K., A.G., and M.V. acknowledge funding from DFF Sapere Aude---EXTREMES project (2067-00009B).

During the preparation of this work, the authors used Claude (Anthropic) and DeepL to improve the readability and language of the paper. After using these tools, the authors reviewed and edited the content as needed and take full responsibility for the content of the article.

\section*{Declaration of interests}
The authors declare no competing interests.

\bibliographystyle{unsrturl-doi}
\bibliography{references}
\onecolumn
% Supplementary: one column; sections, subsections and subsubsections are float barriers
% (as placeins' [section] option, which would force page breaks in the two-column main text)
\makeatletter
\newcommand\@fb@secFB{\FloatBarrier\gdef\@fb@afterHHook{\@fb@topbarrier\gdef\@fb@afterHHook{}}}
\gdef\@fb@afterHHook{}
\g@addto@macro\@afterheading{\@fb@afterHHook}
\expandafter\renewcommand\expandafter\section\expandafter{\expandafter\@fb@secFB\section}
\expandafter\renewcommand\expandafter\subsection\expandafter{\expandafter\@fb@secFB\subsection}
\expandafter\renewcommand\expandafter\subsubsection\expandafter{\expandafter\@fb@secFB\subsubsection}
\makeatother
% Supplementary numbering: Fig. S1, Table S1, Eq. (S1), Sec. S1
\setcounter{figure}{0}
\setcounter{table}{0}
\setcounter{equation}{0}
\setcounter{subsection}{0}
\renewcommand{\thefigure}{S\arabic{figure}}
\renewcommand{\thetable}{S\arabic{table}}
\renewcommand{\theequation}{S\arabic{equation}}
\renewcommand{\thesubsection}{S\arabic{subsection}}
% unique hyperref anchors, otherwise links to Fig. S1 land on Fig. 1
\renewcommand{\theHfigure}{S\arabic{figure}}
\renewcommand{\theHtable}{S\arabic{table}}
\renewcommand{\theHequation}{S\arabic{equation}}
\renewcommand{\theHsubsection}{S\arabic{subsection}}
\input{chapters/05_supplementary}

% \section*{Word Count by Chapter}
% \detailtexcount{chapters/01_introduction}
% \detailtexcount{chapters/02_methods}
% \detailtexcount{chapters/03_results}
% \detailtexcount{chapters/04_discussion_conclusion}

\end{document}

%% file: preamble.tex
\usepackage{verbatim}

\usepackage{amsmath}
\usepackage{amsthm}
\usepackage{amssymb}
\usepackage{mathtools}

\usepackage{siunitx}
\DeclareSIUnit{\pp}{pp}
\usepackage[table, dvipsnames]{xcolor}
\usepackage[shortlabels]{enumitem}
\usepackage{booktabs}
\usepackage{longtable}
\usepackage{threeparttable}
\usepackage{chemformula}
\usepackage[ruled]{algorithm2e}
\usepackage[colorlinks=true, linkcolor=black, citecolor=ForestGreen, urlcolor=blue]{hyperref}
\usepackage{xurl} % break long URLs anywhere (narrow columns)
\usepackage[capitalise, noabbrev]{cleveref}
\usepackage{authblk}
\usepackage{doi}
 
\crefformat{footnote}{#2\footnotemark[#1]#3}

\usepackage[symbol]{footmisc}

\usepackage[UKenglish]{babel}
\usepackage{csquotes}
\usepackage[shortcuts]{extdash}
\usepackage[useregional]{datetime2}
 
\usepackage{graphicx}
\graphicspath{{figures/}}

\usepackage{parskip}
\usepackage{microtype}
\usepackage[left=2cm, right=2cm, top=2.5cm, bottom=2.5cm]{geometry}
\usepackage[small]{titlesec}
\usepackage{float} % keep floats within section
 
\usepackage{stix2}   
\usepackage[font=small, labelfont=bf]{caption}
\usepackage{subcaption}
\usepackage{eurosym}
\usepackage[square,numbers,comma,sort&compress]{natbib}

\usepackage{placeins}
\usepackage{stfloats}
\usepackage{pdflscape}
\usepackage{pgfgantt}
\setlist[itemize]{topsep=4pt,itemsep=2pt}
\definecolor{paperblue}{RGB}{0,102,204}
\definecolor{modelgreen}{RGB}{0,150,100}
\definecolor{deadline}{RGB}{180,0,0}

%% file: chapters/00_graphical_abstract.tex
% Graphical abstract as an unnumbered cover page (preprint only; drop the \input for the journal)
\onecolumn
\pagenumbering{Alph} % own page anchors, so hyperref does not see page 1 twice
\thispagestyle{empty}
\pdfbookmark[1]{Graphical abstract}{graphicalabstract}

% title block set like \maketitle, at the same height as on the title page
% (keep in sync with \title, \author, and \affil in main.tex)
\null\vskip 1em
\begin{center}
  {\LARGE Which wind and solar modelling choices matter? Effects on capacity factors, energy droughts and European power system configuration\par}
  \vskip 1.5em
  {\large L.~Karkossa, A.~Grochowicz, M.~J.~Koivisto, A.~Hahmann, and M.~Victoria\par}
  \vskip 0.5em
  {\small\itshape Department of Wind and Energy Systems, Technical University of Denmark, 2800 Kgs.\ Lyngby, Denmark\par}
\end{center}
\vskip 1.5em

\section*{Graphical abstract}
% trim (left bottom right top): drops the graphic's own heading, which repeats the title above,
% and its side margins, so the boxes line up with the text block
\noindent\includegraphics[width=\textwidth, trim=14.5pt 3pt 14.5pt 28pt, clip]{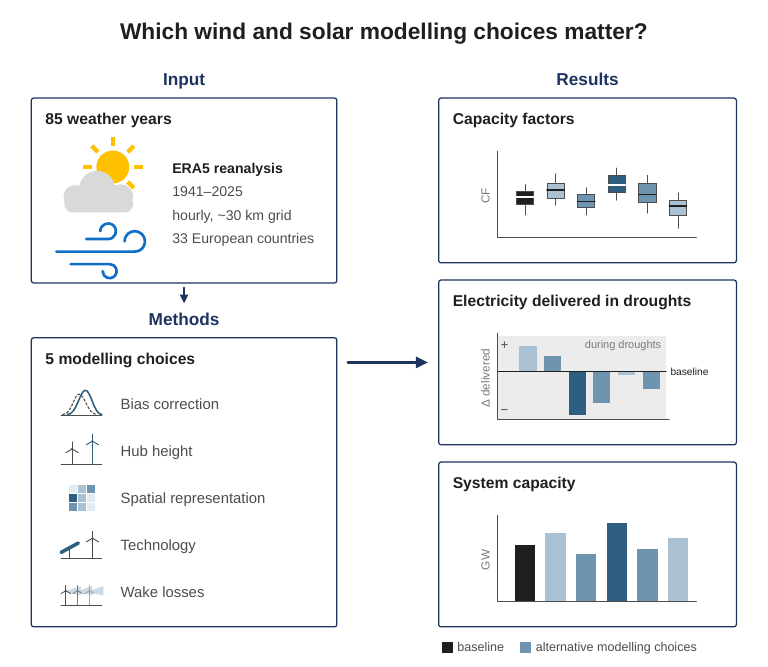}
% note at the foot of the page, set like a footnote
\vfill
\footnoterule
{\footnotesize\noindent The full article starts on the next page.\par}
\clearpage
\pagenumbering{arabic}

%% file: chapters/00_abstract.tex
\begin{abstract}
Europe's future power system will rely heavily on wind and solar power and must remain resilient during prolonged periods of low wind and solar generation, known as variable renewable energy (VRE) droughts. To capture such rare events, adequacy assessments increasingly draw on decades of weather data, but the conversion of weather into power is largely treated as fixed. It remains unclear which modelling choices in this conversion matter for representing these droughts, and whether they matter as much as the weather itself. Here we systematically compare these choices side by side (bias correction, wind speed extrapolation, capacity layouts, turbine and photovoltaic configurations, and wake losses) and trace their effects through VRE droughts to the design of a net-zero European power system, using 85 years of reanalysis data. We find that modelling choices barely change how often VRE droughts occur or when the system is stressed. They do, however, determine how much capacity is needed to withstand that stress. By altering the electricity delivered during droughts, they shift required capacity by up to \qty{480}{\giga\watt}, which is comparable to the spread across the most critical weather years. Modelling exercises should therefore use multiple weather years; bias-correct and regionally validate weather inputs; explicitly represent resource heterogeneity, forward-looking technologies, and wake losses; and judge these choices by their effect on the low-generation tail, not only on the mean.
\end{abstract}

%% file: chapters/01_introduction.tex
\section{Introduction}
The scaling up of renewable deployment \cite{international_energy_agency_renewables_2025, international_renewable_energy_agency_renewable_2026} requires an accurate representation of wind and solar resources within energy system models. 
In highly renewable energy systems, backup capacity requirements are determined by a few meteorological stress events, so-called variable renewable energy (VRE) droughts, rather than by average operating conditions \cite{raynaud_energy_2018, van_der_wiel_influence_2019, ruhnau_storage_2022, van_der_most_extreme_2022, grochowicz_using_2024, gotske_designing_2024, lin_flash_2026, kittel_long-duration_2026, grochowicz_resilience_2026}. 
Weather-related stress is already evident in high electricity prices during winter \textit{Dunkelflauten} or summer heat waves \cite{bundesnetzagentur_results_2025, ember_heat_2025}. 
To support the planning of a resilient European power system, adequacy assessments (e.g.\ ENTSO-E's ERAA \cite{entso-e_european_2025}) rely mainly on reanalyses and, increasingly, climate models. The outcomes of such assessments, therefore, depend on how these weather datasets are converted into renewable generation time series.

A growing body of literature characterises VRE droughts across Europe, highlighting that the frequency, duration, and severity of droughts depend on the identification method \cite{raynaud_energy_2018, otero_copula-based_2022, allen_standardised_2023, kay_variability_2023, kittel_measuring_2024, biewald_evaluation_2025, kittel_multi-threshold_2026}, and examines their meteorological drivers \cite{van_der_wiel_influence_2019,van_der_wiel_meteorological_2019, mockert_meteorological_2023, van_duinen_meteorological_2025, rouges_link_2025}. 
As short climate records poorly capture inter-annual weather variability and low-generation periods \cite{wohland_significant_2019, raynaud_energy_2018, ohlendorf_frequency_2020}, adequacy assessments and energy planning increasingly use multi-decadal weather datasets \cite{pfenninger_dealing_2017, collins_impacts_2018, grochowicz_intersecting_2023, gotske_designing_2024, ullmark_representing_2025, grochowicz_resilience_2026}.

In contrast to weather inputs, their conversion into power has received far less attention, although even small differences in estimated renewable generation can alter adequacy outcomes.

Large-scale energy system studies show sensitivity to bias correction of reanalysis data \cite{andresen_validation_2015,staffell_using_2016,pfenninger_long-term_2016, victoria_using_2019, koivisto_application_2021, gruber_towards_2022, wang_bias_2023}, finer spatial representation of VRE potentials \cite{frysztacki_strong_2021, mattsson_autopilot_2021,jacobson_quantifying_2024, phillips_renewable-homogeneity-based_2026}, updated and emerging technologies \cite{swisher_competitiveness_2022, hodel_which_2024,rahdan_strategic_2025, amaro_e_silva_photovoltaic_2026}, capacity-layout proxies \cite{becker_features_2014}, wake losses \cite{mckenna_high-resolution_2022,borgers_mesoscale_2024, borgers_energy_2025, bramstoft_role_2026},
and atmospheric model-level wind speeds rather than extrapolated ones \cite{hahmann_current_2022, wohland_extrapolation_2024}. 
However, existing studies vary only subsets of these choices, and none evaluates them jointly across capacity factors, energy droughts, and system outcomes (Fig.~\ref{fig:gap_matrix}).

This gap matters particularly for VRE droughts, whose characteristics depend on how weather is converted into power \cite{jung_quantifying_2026,ostraszewski_renewable_2027}. It also reflects the broader disconnect between weather and energy modelling \cite{craig_overcoming_2022}.
Understanding how assumptions in renewable modelling shape these droughts and adequacy challenges is therefore essential for resilient system design.

It remains unclear how the impact of renewable modelling choices compares with inter-annual weather variability; which choices most affect annual capacity factors (CF), VRE drought characteristics, and power system outcomes; and whether their ranking holds across these metrics. 
To address these questions, we conduct a \emph{one-at-a-time} sensitivity analysis of renewable modelling choices using 85 years (1941--2025) of ERA5 reanalysis data for 33 European countries, and rank their influence on capacity factors, VRE drought characteristics, and optimal system design in a capacity and dispatch optimisation model for five critical weather years.

%% file: chapters/02_methods.tex
\section{Methods}

\subsection{Study design and baseline configuration}
\label{sec:baseline}

Hourly solar photovoltaic (PV), onshore and offshore wind generation time series are produced for 33 European countries for the period 1941--2025 using the atlite toolbox v0.4.1 \cite{hofmann_atlite_2021} and ERA5 reanalysis data at \qty{0.25}{\degree} $\times$ \qty{0.25}{\degree} spatial grid spacing \cite{c3s_era5_2018}, with a focus on Denmark, Germany, Italy, and Spain.

\begin{figure*}[tbp]
    \centering
    \includegraphics[width=0.89\linewidth]{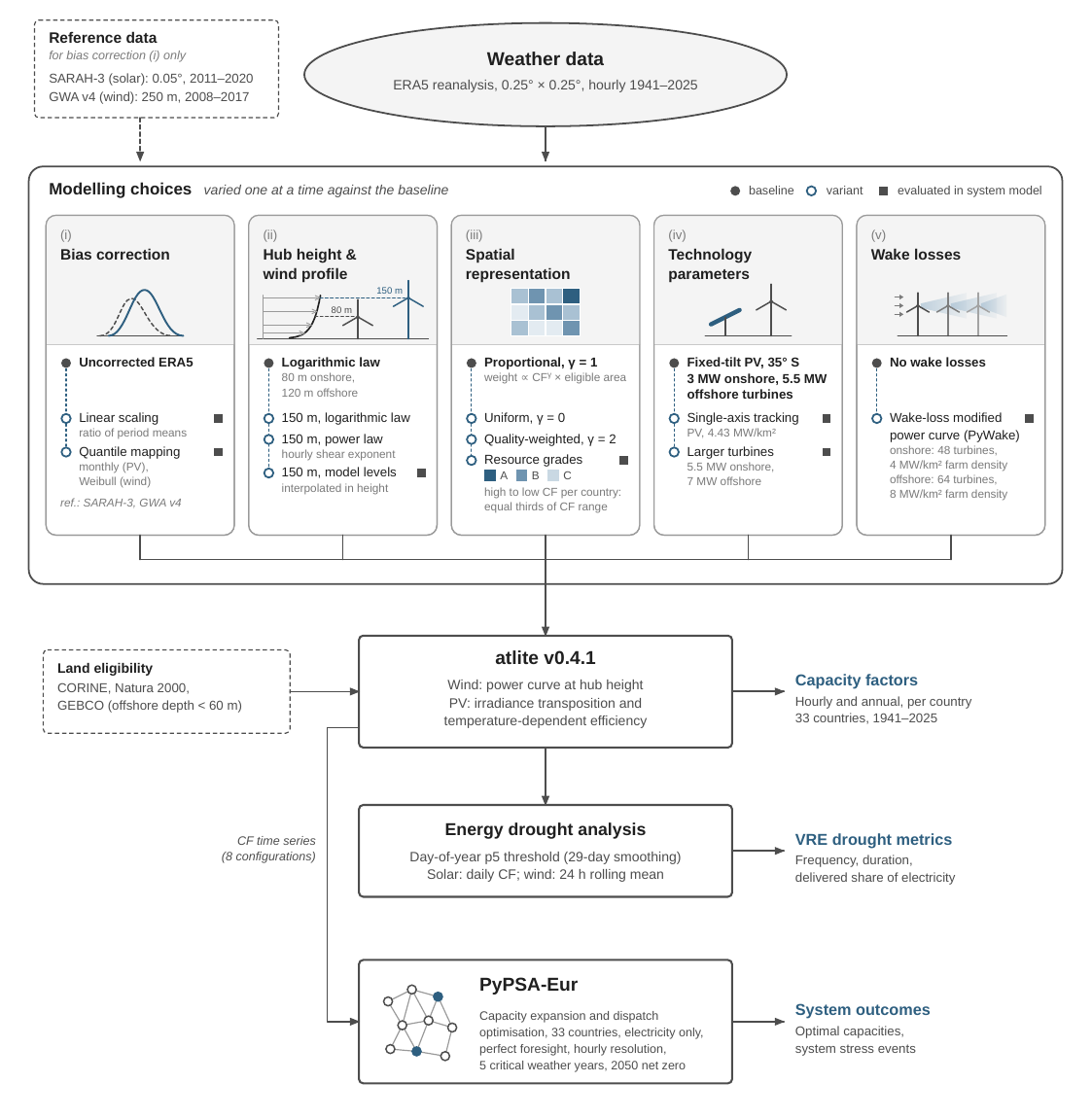}
    \caption{Overview of the modelling chain. Five modelling choices (i)--(v) are varied one at a time against the baseline (filled circles); open circles mark the variants, and squares mark those also evaluated in the system model. Each configuration is converted with atlite into hourly country-level generation time series for 1941--2025 and evaluated at three levels (bottom): (1) mean annual and winter capacity factors; (2) VRE droughts; (3) for the baseline and the variants marked with squares, optimal capacities and system stress events in PyPSA-Eur for five critical weather years. All metrics are reported relative to the baseline. Eligible land excludes protected areas (Natura 2000 \cite{european_environment_agency_eea_natura_2022}) and is determined from CORINE land cover \cite{european_environment_agency_eea_corine_2022}; offshore wind is restricted to water depths below \qty{60}{\meter} (GEBCO \cite{weatherall_new_2015}); see Neumann et al.~\cite{neumann_potential_2023} (Supplementary Material S6) for details.}
    \label{fig:methods_overview}
\end{figure*}

Starting from a baseline configuration (Fig.~\ref{fig:methods_overview}; Sec.~\ref{sec:baseline_config}), we vary one modelling choice at a time: (i) the bias-correction method, (ii) the hub height and vertical wind-profile extrapolation scheme, (iii) the spatial representation and capacity layout, (iv) the technology parameters, and (v) the wake losses (Table~\ref{tab:scenarios}; Secs.~\ref{sec:bias}--\ref{sec:wake-losses}).

Wind power curves are used to convert wind speed at hub height to wind power. Direct and diffuse irradiance on the ground are transposed to calculate global irradiance reaching the surface of solar PV modules, and a temperature-dependent efficiency model is used to estimate solar power time series \cite{hofmann_atlite_2021}.

Modelled CFs are compared to historical  CFs collected by  Nayak et al.~\cite{nayak_validation_2025} (wind, 2015--2022) and IRENA \cite{victoria_solar_2026} (solar, 2020--2024). Note that these historical values are uncertain (Sec.~\ref{sec:limitations}) and are shown for context only.

\subsection{Bias correction}
\label{sec:bias}

Solar irradiance is corrected using the satellite dataset SARAH-3 as the reference \cite{pfeifroth_surface_2023}, which has a higher spatial resolution ($0.05^{\circ} \times 0.05^{\circ}$) and represents cloud cover more accurately than ERA5 \cite{victoria_using_2019, pfeifroth_sarah-3_2024}. Direct and diffuse irradiance are bias-corrected separately. The \qty{100}{m} wind speed is corrected using Global Wind Atlas (GWA) v4 \cite{davis_global_2023}, as ERA5 corrected with an earlier GWA version reproduces large-scale wind generation better than raw ERA5 \cite{murcia_validation_2022,nayak_validation_2025}. Both references are re-gridded to the ERA5 grid and the bias is estimated over ten years.

The \emph{linear correction} rescales each grid cell $i$ by a constant factor $\beta_i$, the ratio of the reference to the ERA5 mean. \emph{Quantile mapping} matches the full distribution, including both tails, to the reference \cite{cannon_bias_2015}. To account for seasonality, we use monthly empirical distributions for solar. Since GWA provides only Weibull parameters, a Weibull-to-Weibull transfer function is used for wind. Neither method distorts the diurnal or seasonal cycles (Figs.~\ref{fig:diurnal_cycle_solar}--\ref{fig:seasonal_cycle_onwind}). The formulation, reference periods, and correction maps are given in Sec.~\ref{sec:supp_bias}. Correction factors and transfer functions are available on  Zenodo~\cite{karkossa_bias-correction_2026}.

\begin{table*}[tbp]
\centering
\small
\caption{Modelling configurations considered. Each configuration differs from the baseline in exactly one assumption. The technology columns indicate which generation time series are affected. The last column indicates whether the configuration is additionally evaluated in the capacity expansion model (Sec.~\ref{sec:energy_system_analysis_methods}). All configurations are included in the annual mean capacity-factor and VRE drought analyses.}
\label{tab:scenarios}
\begin{threeparttable}
\begin{tabular}{@{}llcccc@{}}
\toprule
Configuration & Change relative to the baseline & PV & Onshore wind & Offshore wind & System \\
\midrule
Baseline & Sec.~\ref{sec:baseline} & \checkmark & \checkmark & \checkmark & \checkmark \\
\addlinespace
\multicolumn{6}{@{}l}{\emph{Bias correction} (Sec.~\ref{sec:bias})}\\
Linear correction  & Scaling by the ratio of period means & \checkmark & \checkmark & \checkmark & \checkmark \\
Quantile mapping   & Monthly empirical (solar), Weibull (wind) & \checkmark & \checkmark & \checkmark & \checkmark \\
\addlinespace
\multicolumn{6}{@{}l}{\emph{Vertical wind profile and hub height} (Sec.~\ref{sec:wind_height})\tnote{a}}\\
\qty{150}{\meter} logarithmic & Log-law extrapolation to \qty{150}{m} &  & \checkmark & \checkmark &  \\
\qty{150}{\meter} power law   & Hourly shear exponent from 10 and \qty{100}{m} &  & \checkmark & \checkmark &  \\
\qty{150}{\meter} model level & Interpolation of ERA5 model-level winds &  & \checkmark & \checkmark & \checkmark \\
\addlinespace
\multicolumn{6}{@{}l}{\emph{Spatial representation} (Sec.~\ref{sec:spatial})}\\
Uniform layout          & Layout exponent $\gamma = 0$ & \checkmark & \checkmark & \checkmark &  \\
Quality-weighted layout & Layout exponent $\gamma = 2$ & \checkmark & \checkmark & \checkmark &  \\
Resource Grade~A & High-quality grid cells only\tnote{b}   & \checkmark & \checkmark & \checkmark & \checkmark \\
Resource Grade~B & Medium-quality grid cells only\tnote{b} & \checkmark & \checkmark & \checkmark & \checkmark \\
Resource Grade~C & Low-quality grid cells only\tnote{b}    & \checkmark & \checkmark & \checkmark & \checkmark \\
\addlinespace
\multicolumn{6}{@{}l}{\emph{Technology parameters} (Sec.~\ref{sec:tech})}\\
Solar HSAT     & Horizontal single-axis tracking replaces fixed tilt & \checkmark &  &  & \checkmark \\
Larger turbine & Higher-capacity turbine at baseline hub height &  & \checkmark & \checkmark & \checkmark \\
\addlinespace
\multicolumn{6}{@{}l}{\emph{Wake losses} (Sec.~\ref{sec:wake-losses})}\\
Wake losses & Wake-modified power curve &  & \checkmark & \checkmark & \checkmark \\
\bottomrule
\end{tabular}
\begin{tablenotes}[flushleft]
\footnotesize
\item[a] All three share a new hub height of \qty{150}{m}. Comparisons among them isolate the extrapolation scheme. The comparison with the baseline also includes the effect of the change in hub height from \qty{80}{\meter} (onshore) and \qty{120}{\meter} (offshore) to \qty{150}{m}. Only the model-level scheme is carried into the capacity expansion model, where this configuration is intended as a hub-height sensitivity rather than a comparison of extrapolation schemes.
\item[b] In the capacity-factor and VRE drought analyses, each grade is evaluated as a separate time series. In the system model, the three grades of a country coexist as separate generators, each with its own profile and installable potential, so the optimisation chooses among them.
\end{tablenotes}
\end{threeparttable}
\end{table*}

\subsection{Wind speed at hub height}
\label{sec:wind_height}
Hub-height wind speed is derived in three ways: (i) logarithmic extrapolation of the \qty{100}{m} wind speed with the hourly ERA5 roughness length under neutral stratification \cite{andresen_validation_2015}, as in the baseline; (ii) a power law with an hourly shear exponent from the \qty{10}{m} and \qty{100}{m} wind speeds; and (iii) linear interpolation between ERA5 model levels, whose heights are reconstructed hourly via the hypsometric equation \cite{hahmann_current_2022}. A detailed description is given in Sec.~\ref{sec:supp_wind_height}. All three are applied at a common hub height of \qty{150}{m} for onshore and offshore wind, so that comparisons among them isolate the extrapolation scheme.

\subsection{Spatial representation and resource grades}
\label{sec:spatial}
Using the full 1941--2025 time series, we compute mean capacity factors $\overline{CF}_i$ for all grid cells. Capacity layouts are then constructed by assigning a weight $w_i$ to each cell $i$,
\begin{equation}
w_i = \left(\overline{CF}_i\right)^{\gamma} S_i \, \kappa_i^{\max},
\end{equation}
where $S_i$ is the eligible area of the cell and $\kappa_i^{\max}$ the maximum installable capacity per unit area given in Sec.~\ref{sec:baseline_config}. These weights distribute a country's installed capacity across its cells and are used to aggregate the cell CFs into the country-level time series. The exponent $\gamma$ controls how strongly the layout prioritises high-resource cells: for $\gamma=0$, the capacity is distributed only by available area, a so-called \emph{uniform} layout. The proportional layout with $\gamma=1$ corresponds to the \emph{baseline}. With $\gamma=2$, a \emph{quality-weighted} layout is tested.    

Because a country-level CF averages the best and worst sites, we additionally split each country's range of cell-mean CFs into three equal intervals, forming Resource Grades~A (upper third), B, and C (lower third), fixed across all years (Sec.~\ref{sec:supp_spatial}). The split is by resource quality, not by area, so a country with only a few exceptionally good sites has a small area corresponding to Grade~A.

\subsection{Technology parameters}
\label{sec:tech}

For PV, horizontal single-axis tracking (HSAT) replaces the fixed-tilt baseline. HSAT increases annual yield and broadens the daily production window, which makes it strategically valuable for the energy system \cite{rahdan_strategic_2025}. For wind, a larger turbine with lower specific power replaces the baseline turbine at the same hub height: \qty{5.5}{\mega\watt} (\qty{229}{\watt\per\square\meter}) instead of \qty{3}{\mega\watt} (\qty{305}{\watt\per\square\meter}) onshore, and \qty{7}{\mega\watt} (\qty{223}{\watt\per\square\meter}) instead of \qty{5.5}{\mega\watt} offshore (Sec.~\ref{sec:supp_tech_wake}).

\subsection{Wake losses}
\label{sec:wake-losses}
Wake losses are represented by a wake-modified power curve, computed with PyWake v2.6.2 \cite{simutis_enhancing_2024} for a representative circular wind farm with an installation density of \qty{4}{\mega\watt\per\square\kilo\meter} onshore and \qty{8}{\mega\watt\per\square\kilo\meter} offshore \cite{covey_empirical_2026}. The curve accounts for near-farm and mesoscale losses and is averaged over all wind directions (Sec.~\ref{sec:supp_tech_wake}).

\subsection{Capacity factor and VRE drought metrics}
\label{sec:droughts}
\begin{figure*}[tbp]
    \centering
    \includegraphics[width=0.89\linewidth]{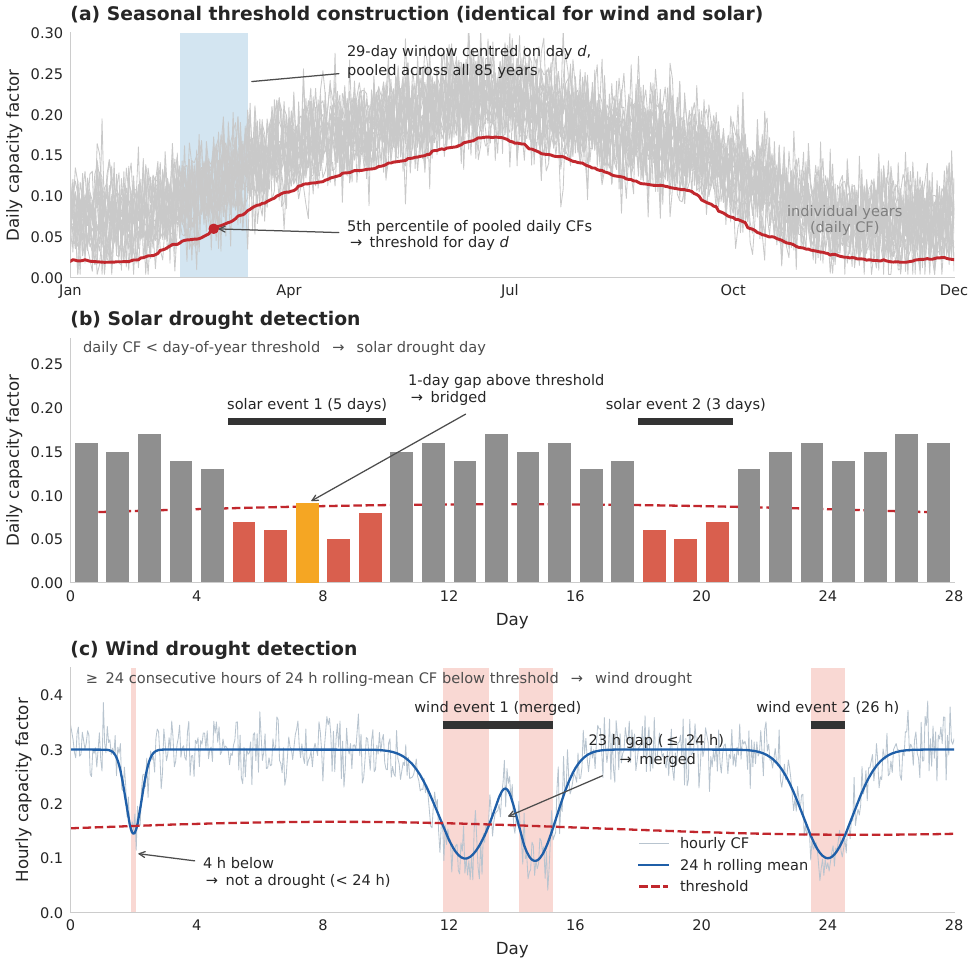}
    \caption{Illustration of the VRE drought identification procedure. (a) Construction of the daily-varying p5 threshold for wind and solar, computed for each day of the year across all 85 years and smoothed with a centred 29-day moving average. (b) Identification of solar drought days and events from daily capacity factors, with a minimum duration of one day and bridging of one-day interruptions above the threshold. (c) Identification of wind droughts from the hourly 24-hour rolling-mean capacity factor. Periods below the threshold must persist for at least 24 consecutive hours, and events separated by no more than 24 hours above the threshold are merged.}
    \label{fig:drought_identification}
\end{figure*}

VRE droughts are identified separately for each country, technology, and configuration (Fig.~\ref{fig:drought_identification}). The detection threshold ${CF}^{\text{thr}}_{c,d}$ is the 5th percentile (p5) of daily CFs for each country $c$ and day of the year $d$ across the 85-year record, smoothed with a centred 29-day moving average to account for the seasonal cycle. Solar droughts are identified from daily CFs, and wind droughts from the 24-hour rolling mean of hourly CFs. As no universal VRE drought definition exists, the identified events depend on threshold definition, temporal and spatial aggregation, minimum-duration criteria, and event-merging rules \cite{kittel_measuring_2024, kittel_multi-threshold_2026}. We use p5 to focus on exceptionally low generation, and thresholds up to the 15th percentile yield the same conclusions (Sec.~\ref{sec:supp_threshold}).

Generation during an event is measured against the seasonal norm ${CF}^{\text{norm}}_{c,d}$, the smoothed day-of-year median CF. With $\overline{CF}_e$ the mean CF during event $e$ and $\overline{CF}^{\text{norm}}_e$ the mean seasonal norm over the same period, the delivered share of electricity during each event is 
\begin{equation}
\delta_e = 100\, \frac{\overline{CF}_e}{\overline{CF}^{\text{norm}}_e} \qquad [\unit{\percent}].
\end{equation}
A value of \qty{100}{\percent} means generation matches the seasonal norm. Lower values indicate more severe droughts. Aggregated over events, 
\begin{equation}
\bar{\delta}_c = 100\, \frac{\sum_e L_e\, \overline{CF}_e}{\sum_e L_e\, \overline{CF}^{\text{norm}}_e} \qquad [\unit{\percent}],
\label{eq:delivered}
\end{equation}
where the sums run over all events of a country, technology, and season, and $L_e$ is the event duration. This weights each $\delta_e$ by the event's expected generation $L_e\, \overline{CF}^{\text{norm}}_e$, so that longer droughts, which cause larger supply deficits, carry more weight. As ${CF}^{\text{thr}}_{c,d}$ and ${CF}^{\text{norm}}_{c,d}$ derive from each configuration's own CFs, $\bar{\delta}_c$ is relative to that configuration's own climatology. We report event frequency, mean duration, and $\bar{\delta}_c$, the latter in percentage points relative to the baseline.

\subsection{System analysis}
\label{sec:energy_system_analysis_methods}
The eight configurations marked in the System column of Table~\ref{tab:scenarios} (the three resource grades form one) are each used in a perfect-foresight capacity expansion and dispatch optimisation with PyPSA-Eur, closely following Grochowicz et al.~\cite{grochowicz_using_2024}. 
We consider an electricity-only system covering 33 countries, with hourly resolution for a full year. Electricity can be generated by onshore and offshore wind, solar PV, hydropower, nuclear, H\textsubscript{2} turbines, and biomass generators, and can be stored in batteries, hydrogen storage, and pumped hydro storage (PHS).
The capacities of all generators and storage can be expanded, except for biomass, hydropower, and PHS, which are fixed at 2023 values. Nuclear capacities from the same year are used as a lower bound.
We evaluate five critical weather years (1941/42, 1962/63, 1965/66, 1996/97, and 2010/11), identified by Grochowicz et al.~\cite{grochowicz_using_2024}. The analysis is conducted for a net-zero scenario without carbon capture, using 2023 transmission capacities. Costs follow technology-data v0.13.2 (2050)~\cite{lisazeyen_pypsatechnology-data_2025}, which distinguishes HSAT but not turbine types or hub heights, so the turbine and hub-height configurations retain baseline turbine costs (Sec.~\ref{sec:limitations}).

System stress events are identified from the dual variables of the nodal energy-balance constraints (i.e.\ electricity prices), which are load-weighted and averaged into a single representative hourly European electricity price. Within each weather year, we identify periods where this price remains above that year's 97th percentile for at least 24 hours, extend them outwards along the steep ramps into and out of the event (see e.g.\ Fig.~\ref{fig:stress_event_1}), and retain the period with the largest accumulated price above the threshold as that year's stress event. Detection is run once on the baseline networks, with windows fixed across all configurations, and once independently per configuration to test whether modelling choices shift the timing of the most severe periods (Figs.~\ref{fig:stress_events_overview}--\ref{fig:stress_event_5}).

%% file: chapters/03_results.tex
\section{Results}
\subsection{Average annual country capacity factors}
\begin{figure*}[tbp]
  \centering
  \includegraphics[width=0.89\linewidth]{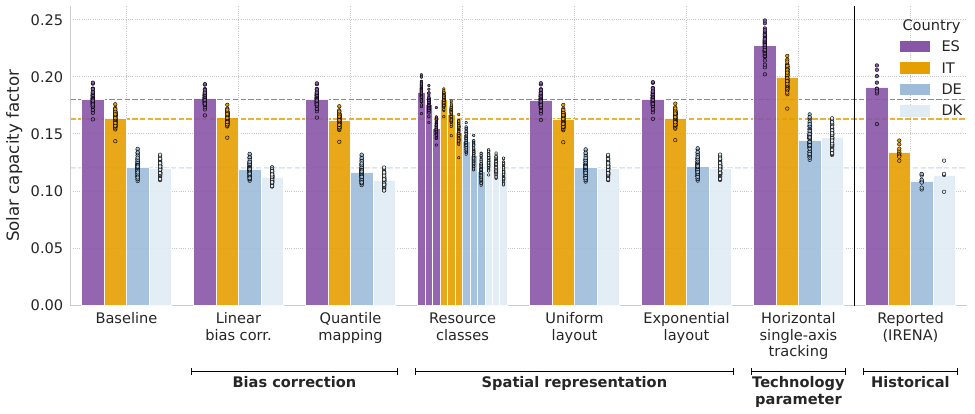}
  \caption{Mean annual solar capacity factors across modelling assumptions in Spain (ES), Italy (IT), Germany (DE), and Denmark (DK). Bars indicate the mean over the 85 weather years, dots the individual annual values, and dashed lines the baseline mean of each country. For resource grades, the three bars per country show Grades~A to C from left to right. Configurations are grouped into bias correction, spatial representation, and technology parameters. Historical values derived from IRENA statistics are shown for context only, as the modelled PV fleet is not matched to the historical fleet. They show no consistent deviation from the baseline (ranging from \qty{-18}{\percent} (\qty{-2.9}{\pp}) in Italy to \qty{+6}{\percent} (\qty{+1.1}{\pp}) in Spain) and should be treated with care: reported CFs refer to DC capacities, whereas PV modules are commonly overbuilt relative to inverter and grid-connection capacities, which lowers the estimated CF.}

  \label{fig:solar-cf-yearly}
\end{figure*}

\begin{figure*}[tbp]
  \centering
  \includegraphics[width=0.89\linewidth]{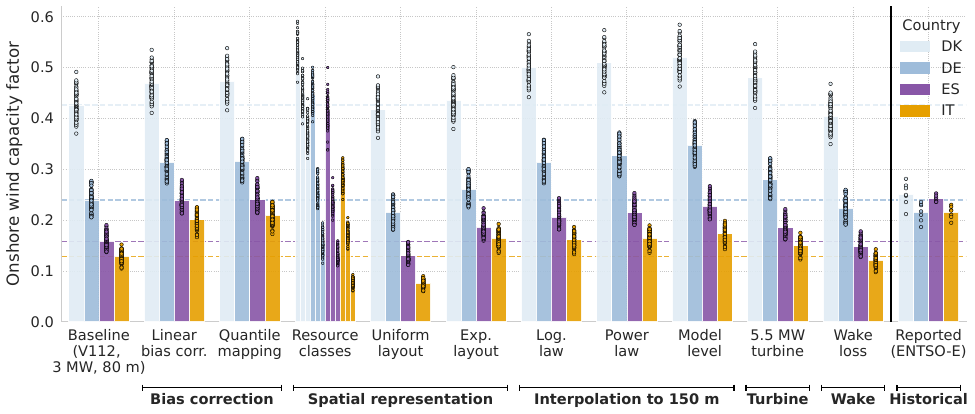}
  \caption{Mean annual onshore wind capacity factors across modelling assumptions in Denmark (DK), Germany (DE), Spain (ES), and Italy (IT). Bars show the mean over the 85 weather years, dots the individual annual values, and dashed lines the baseline mean of each country. For resource grades, the three bars per country show Grades~A to C from left to right. Configurations are grouped into bias correction, spatial representation, hub-height interpolation methods at \qty{150}{\metre}, turbine type, and wake losses. Historical values for 2015--2022 from Nayak et al.~\cite{nayak_validation_2025} are shown for context only, as the modelled capacity layout and turbines are not matched to the historical fleet.}
  \label{fig:onshore-cf-yearly}
\end{figure*}

\begin{figure*}[b]
  \centering
  \includegraphics[width=0.8\linewidth]{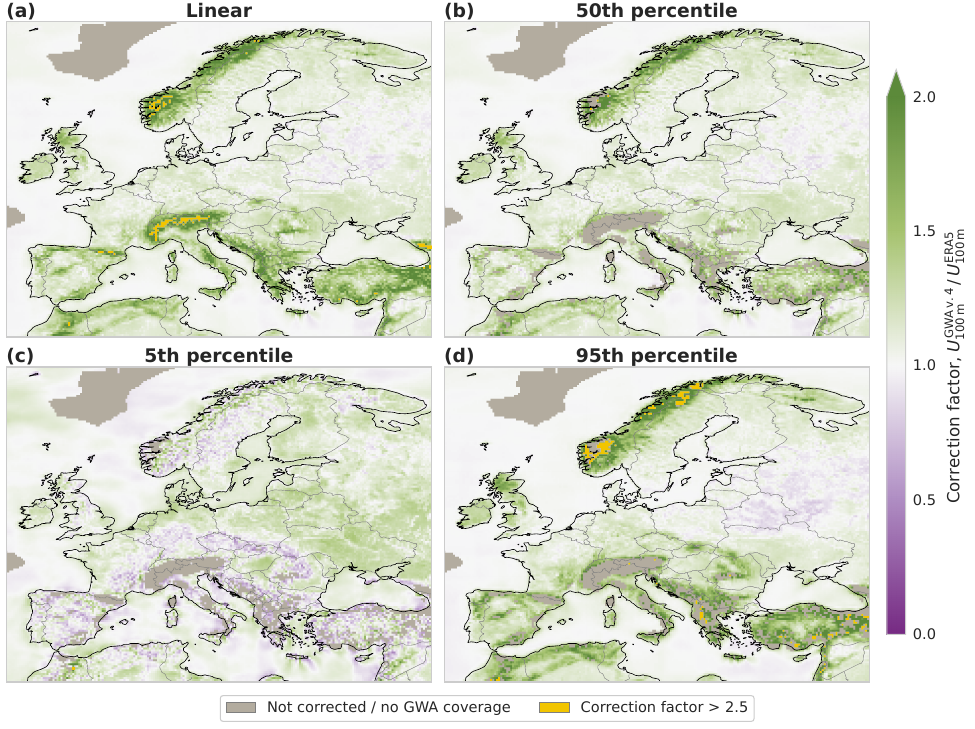}
  \caption{Spatial distribution of the \qty{100}{\meter} wind-speed bias correction factors derived from the Global Wind Atlas (GWA) v4 and ERA5 for the 2008--2017 reference period. Panel (a) shows the linear correction factor, while panels (b)--(d) show quantile-mapping correction factors for the 50th, 5th, and 95th percentiles, respectively. Values greater than one indicate upward corrections of ERA5 wind speeds. Grey areas denote regions without corrections or GWA v4 coverage, while yellow areas indicate correction factors exceeding 2.5.}
  \label{fig:bias_correction_era5_gwa4_vertical}
\end{figure*}

Across the 85-year hourly time series of solar PV, horizontal single-axis tracking has the largest impact on solar CF among the modelling assumptions, increasing it in all four countries by \qtyrange{19}{26}{\percent} (\numrange{2.3}{4.7}~percentage points (pp); Fig.~\ref{fig:solar-cf-yearly}). This change in the temporal profile, which shifts generation into the morning and evening hours  (Fig.~\ref{fig:diurnal_cycle_solar}), outweighs the effect of spatial heterogeneity captured by the resource grades (\qtyrange{+4}{+19}{\percent} for high- and \qtyrange{-3}{-14}{\percent} for low-resource grades), whereas the spatial layout of the capacity does not affect the mean CF. The choice of weather year can be of similar importance: individual years deviate from the 85-year mean by up to \qty{-12}{\percent} (\qty{-2.0}{\pp}) and \qty{+13}{\percent} (\qty{+1.6}{\pp}), which exceeds the effects of layout assumptions and, in most countries, of bias correction. Bias correction has a negligible impact in Spain and Italy, a slight one in Germany (\qty{-3.6}{\percent}, \qty{-0.4}{\pp}), and a substantial reduction in Denmark (\qty{-8.9}{\percent}, \qty{-1.1}{\pp}). In winter, when generation is low, quantile mapping reduces the Danish CF by up to \qty{28}{\percent} (\qty{-1.2}{\pp}; Fig.~\ref{fig:solar-cf-winter}). %Historical (IRENA) values show no consistent deviation from the baseline, ranging from \qty{-18}{\percent} (\qty{-2.9}{\pp}) in Italy to \qty{+6}{\percent} (\qty{+1.1}{\pp}) in Spain, and should be treated with care: reported CFs refer to nominal ratings, whilst PV plants are commonly overbuilt relative to their inverter and grid-connection capacities, which lowers the realised CF.

Onshore wind shows greater variation in the average CF across countries and configurations than solar PV (Fig.~\ref{fig:onshore-cf-yearly}). Inter-annual variability is relatively high, with deviations of up to \qty{6.5}{\pp} over the 85 years, yet the representation of the resource within a country can matter even more: high-resource sites raise mean CFs by \qtyrange{11}{26}{\pp}, and bias correction increases them by about \qty{8}{\pp} in countries with complex terrain such as Spain and Italy, exceeding several technology effects. Underestimation of wind velocity over Southern Europe in ERA5 has been reported previously \cite{murcia_validation_2022}, and this upward correction, regardless of whether linear correction or quantile mapping is used, brings annual values closer to historical data. Raising the hub height to \qty{150}{\metre} (increasing CF by \qtyrange{3}{11}{\pp}) has a larger effect than switching to the larger turbine type (\qtyrange{+2.1}{+5.3}{\pp}) or including wake losses (\qtyrange{-0.8}{-2.2}{\pp}), while the method used to derive hub-height wind speeds is less important.

While bias correction narrows the gap between modelled and historical CFs in Spain and Italy, it lifts the Danish and German CFs, which the baseline already overestimates, further away from the reported values. Historical values are subject to uncertainty, as alternative sources report different installed capacities \cite{nayak_validation_2025}, and reported generation can be affected by curtailment. Nonetheless, bias correction with GWA v2 instead of v4 yields CFs closer to the reported values in Denmark and Germany \cite{murcia_validation_2022}; the difference between the two versions is shown in Figs.~\ref{fig:bias_factor_gwa2_gwa4_era5} and~\ref{fig:bias_factor_gwa2_gwa4}. Offshore wind behaves similarly to onshore wind, but with less variability between modelling choices (Sec.~\ref{sec:supp_offshore_cf}).

\subsection{Periods of low VRE generation}

\begin{figure*}[tbp]
    \centering
    \includegraphics[width=0.71\linewidth]{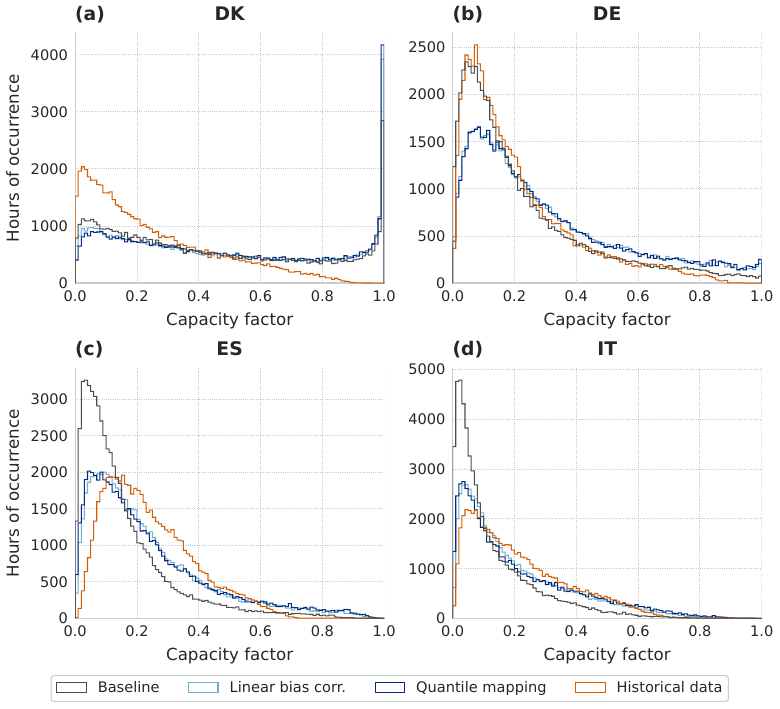}
    \caption{Hourly onshore wind capacity factor distribution for the years 2015--2022 for (a) Denmark, (b) Germany, (c) Spain, and (d) Italy, comparing the modelled baseline with linear correction, quantile mapping, and historical data~\cite{nayak_validation_2025}.}
    \label{fig:cf_distribution_with_reference}
\end{figure*}

Low wind generation is a major driver of system stress in Europe \cite{van_der_wiel_influence_2019, meng_climate-driven_2025}. Quantile mapping corrects the low and upper tails of the wind speed distribution differently (Fig.~\ref{fig:bias_correction_era5_gwa4_vertical}c,d). In some regions, e.g.\ Spain and Italy, it amplifies the extremes: low wind speeds are corrected downward, while high wind speeds are corrected upward, thereby widening the distribution. Linear correction (Fig.~\ref{fig:bias_correction_era5_gwa4_vertical}a), which closely resembles the median correction of quantile mapping, cannot reproduce this tail behaviour. Yet after spatial aggregation and conversion to power, both bias correction techniques yield similar CF distributions, with tails varying only slightly (Fig.~\ref{fig:cf_distribution_with_reference}; Sec.~\ref{sec:supp_bias_cf_dist}).

\begin{figure*}[tbp]
    \centering
    \includegraphics[width=0.89\linewidth]{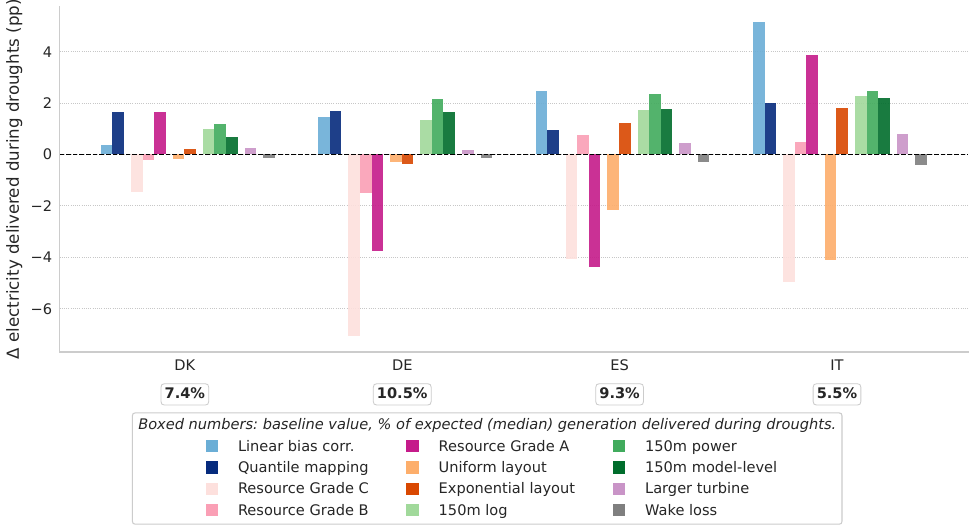}
    \caption{Change in the percentage of expected (day-of-year median for 85 years) electricity delivered during wind-drought periods (below each configuration's p5 threshold), relative to the onshore-wind baseline, for (a) Denmark, (b) Germany, (c) Spain, and (d) Italy. Values are shown for the winter season, in percentage points; positive values indicate a higher share of electricity delivered during a drought than the baseline. Numbers below the x-axis represent the baseline value for reference: the percentage of expected electricity delivered during droughts.}
    \label{fig:drought_delivered_onwind_winter_countries}
\end{figure*}

The frequency and duration of identified low-wind periods (Sec.~\ref{sec:droughts}) remain largely unchanged across modelling choices, deviating from the baseline by only about \qty{2.4}{\percent} on average (Sec.~\ref{sec:supp_drought_onwind}). This indicates that the occurrence of drought events is relatively robust to the assumptions investigated. Their severity, however, measured as the share of expected electricity delivered during the droughts (Eq.~\eqref{eq:delivered}), shifts in opposite directions depending on the modelling choice (Fig.~\ref{fig:drought_delivered_onwind_winter_countries}). Bias correction and higher hub heights increase the delivered share in all four countries, thereby reducing the shortfall, and do so in every season of the load-weighted European average (Figs.~\ref{fig:drought_delivered_onwind_seasons_countries} and~\ref{fig:drought_delivered_onwind_seasons_europe_avg}). Despite their very similar CF distributions, linear correction increases the delivered share more than quantile mapping in complex terrain such as Spain and Italy. Resource grades show no uniform direction: Resource Grade~A delivers more in Denmark and Italy. Resource Grade~C has by far the largest single effect, reducing the delivered share by up to \qty{7}{\pp} in Germany.

Wake losses consistently deepen the shortfall, also at the European level, but, like turbine size and capacity layout, they have a comparatively small effect. Offshore wind broadly follows the same pattern, with smaller effects of hub height, capacity layout, and turbine size, but larger reductions from Resource Grade~C and wake losses (Figs.~\ref{fig:drought_delivered_offwind-ac_seasons_countries} and~\ref{fig:drought_delivered_offwind-ac_seasons_europe_avg}).

For solar (Fig.~\ref{fig:drought_delivered_solar_seasons_europe_avg}), the pattern is similar. Bias correction (again, linear more than quantile mapping) increases the delivered share, while all resource grades and HSAT reduce it. The effect of changing the capacity layouts is negligible. At the country level, however, the picture is more diverse (Fig.~\ref{fig:drought_delivered_solar_seasons_countries}).

\subsection{Different renewable modelling assumptions change the system configuration}
\label{sec:results_system_configuration}
\begin{figure*}[tbp]
  \centering
  \includegraphics[width=0.89\linewidth]{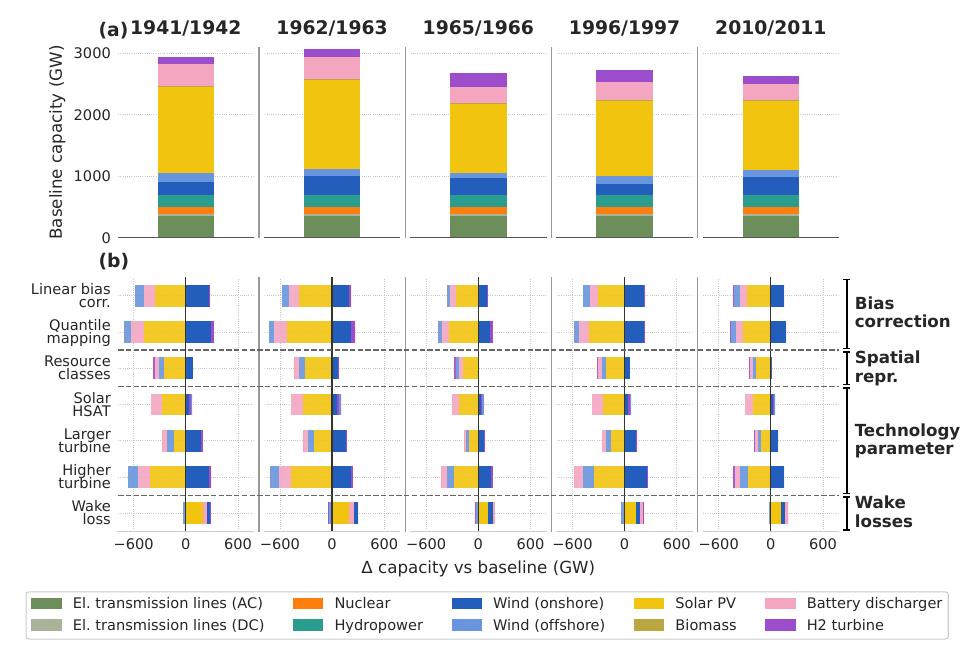}
  \caption{Total installed system capacity per weather year. (a) Baseline capacity by technology for each of the five weather years. (b) Change in system capacity relative to the corresponding baseline for each modelling assumption. Positive values indicate additional capacity compared to the baseline, while negative values indicate reduced capacity.}
  \label{fig:capacity_comparison_system}
\end{figure*}

We measure the sensitivity of power system planning to VRE modelling assumptions by co-optimising capacity and dispatch independently for each of the five critical weather years \cite{grochowicz_using_2024} (Sec.~\ref{sec:energy_system_analysis_methods}). The inter-annual variability of total system capacity in the baseline (\qty{430}{\giga\watt} across the five years (Fig.~\ref{fig:capacity_comparison_system}a)) is of similar magnitude to the largest changes induced by modelling assumptions (\qtyrange{295}{480}{\giga\watt} per weather year (Fig.~\ref{fig:capacity_comparison_system}b)). The picture differs by technology, however: backup capacities (such as H\textsubscript{2} turbines) are driven more strongly by the weather year than by any modelling assumption.

Under both methods, bias correction raises the load-weighted European average onshore wind CF by about \qtyrange{12}{13}{\percent} annually and \qty{29}{\percent} in winter (Fig.~\ref{fig:onshore-cf-winter}), and lifts the lower tail of its distribution, making onshore wind more competitive. Across weather years, onshore wind (\qtyrange{+95}{+290}{\giga\watt}) therefore replaces solar PV (\qtyrange{-250}{-520}{\giga\watt}) and offshore wind (\qtyrange{-40}{-100}{\giga\watt}), reducing total system capacity by \qtyrange{9}{15}{\percent}. As the onshore wind CF is double that of solar PV, the substitution is far from one-to-one in terms of capacity. With less solar generation to shift within the day, battery discharge falls by \qtyrange{27}{42}{\percent} (Fig.~\ref{fig:energy_supply_system}). These effects are more pronounced under quantile mapping, which lowers the solar winter CF considerably more than linear correction (load-weighted European average: \qty{-12}{\percent} versus \qty{-3}{\percent}; Fig.~\ref{fig:solar-cf-winter}). It therefore displaces more solar PV (\num{-414} versus \qty{-317}{\giga\watt} on average) and battery capacity (\num{-114} versus \qty{-94}{\giga\watt}, i.e.\ \qty{-37}{\percent} versus \qty{-31}{\percent}), with consequences for backup capacity during stress events (Sec.~\ref{sec:results_stress_events}).

Most refinements in technology representation and spatial resolution increase the annual CF and thus reduce total system capacity: \qty{150}{\meter} hub heights by \qtyrange{10}{16}{\percent}, HSAT and resource grades (offered as separate generators) each by about \qtyrange{9}{12}{\percent}, and larger turbines by \qtyrange{3}{5}{\percent}. Wake losses are the exception. Neglecting them underestimates total system capacity by \qtyrange{6}{8}{\percent}, as they shift capacity from offshore wind (\qtyrange{-24}{-45}{\giga\watt}) to solar PV (\qtyrange{+115}{+195}{\giga\watt}), onshore wind (\qtyrange{+33}{+52}{\giga\watt}), and batteries (\qtyrange{+21}{+53}{\giga\watt}).

\subsection{System stress events require adequate resource representation}
\label{sec:results_stress_events}
\begin{figure*}[b]
  \centering
  \includegraphics[width=0.89\linewidth]{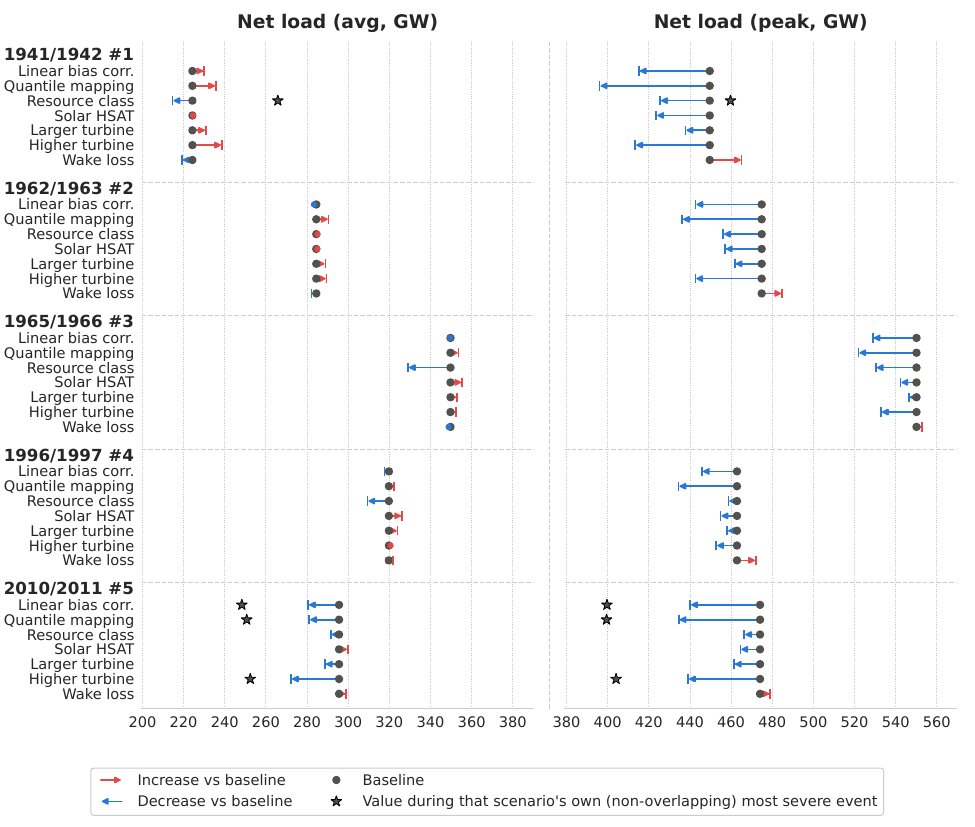}
  \caption{Sensitivity of the average (left) and peak (right) net load during the five system stress events to the renewable-generation modelling assumptions. Grey dots mark the baseline; arrows show each configuration over the same period, the baseline's stress window (red: increase; blue: decrease). In four cases, a configuration's own most severe period falls on different dates (2010/11 under both bias corrections and the \qty{150}{\metre} hub height, and 1941/42 under resource grades; Fig.~\ref{fig:stress_events_overview}); for these, a star additionally shows the value over the configuration's own stress window.}
  \label{fig:system_stress_events}
\end{figure*}

System stress events are periods of acute supply scarcity; in each weather year, we identify the period with the highest accumulated hourly cost above that year's 97th percentile (Sec.~\ref{sec:energy_system_analysis_methods}). The identified events are largely independent of the renewable-generation modelling assumptions in both timing and net load, except in 2010/11, where both bias corrections and the \qty{150}{\metre} hub height identify a different event window, as do the resource grades in 1941/42 (Fig.~\ref{fig:stress_events_overview}). Differences in average and peak net load are driven mostly by inter-annual variability rather than by modelling assumptions (Fig.~\ref{fig:system_stress_events}). Still, peak net load falls by \qtyrange{8}{54}{\giga\watt} under bias correction, HSAT, and \qty{150}{\metre} hub heights; under bias correction, the additional onshore wind capacity delivers \qtyrange{30}{92}{\giga\watt} more during the peak net-load hour.

Counterintuitively, a reduced peak net load may increase the need for backup capacity compared to the baseline: bias correction displaces solar PV, most notably under quantile mapping, limiting the surplus that charges batteries (Sec.~\ref{sec:results_system_configuration}). The resulting battery fleets are \qtyrange{30}{37}{\percent} smaller and, although similarly charged, store less energy when stress periods begin, a shortfall that H\textsubscript{2} turbines must compensate for (Fig.~\ref{fig:stress_battery_h2_daily}). During the peak net-load hour, the drop in battery discharge (linear correction: \qtyrange{-29}{-46}{\giga\watt}; quantile mapping: \qtyrange{-39}{-73}{\giga\watt}) exceeds the reduction in peak net load in all weather years except 2010/11, so H\textsubscript{2}-turbine capacity increases by \qty{7}{\percent} and \qty{16}{\percent} on average, respectively.

Although the duration and frequency of VRE droughts are largely robust across the modelling assumptions considered here, the amount of generation available during these periods is not. These differences propagate into system design, altering the balance between renewable generation, storage, and dispatchable backup capacity.

%% file: chapters/04_discussion_conclusion.tex
\section{Discussion \& Conclusions}
\label{sec:discussion}

\subsection{Comparison with previous work}
\label{sec:key_findings}
Consistent with previous studies \cite{collins_impacts_2018,ruhnau_storage_2022,grochowicz_intersecting_2023,gotske_designing_2024}, inter-annual weather variability across the 85 years remains one of the dominant sources of uncertainty in power system planning. However, the most influential renewable modelling choices are comparable in magnitude. At the system level, the largest modelling-induced changes in installed capacity (\qtyrange{295}{480}{\giga\watt}) match the spread across the five critical weather years (\qty{430}{\giga\watt}). The conversion of weather data into generation time series should therefore be treated as a source of uncertainty comparable to that associated with the choice of weather years.

Technology representation and spatial resolution are known to shape energy system model outcomes \cite{frysztacki_strong_2021,jacobson_quantifying_2024}; we show that the conversion of weather to power does so as well. Mean capacity factors respond most to resource grades, hub height, HSAT, and (for onshore wind in complex terrain) bias correction. The frequency and duration of VRE droughts barely change across modelling choices (mean deviations of \qtyrange{2}{3}{\percent}), and the most severe system stress events largely keep their timing. Drought severity, however, does change, and with it the capacity needed to cover these events. Choices that raise mean capacity factors can even lower the delivered share during droughts: HSAT deepens solar shortfalls in low-irradiance periods, and concentrating wind capacity at high-yield sites (Resource Grade~A) can reduce the spatial diversity that buffers low-generation hours \cite{handschy_is_2017}. Yield-oriented siting has already been questioned on cost and curtailment grounds \cite{radu_assessing_2022,jung_projections_2022}; our results suggest that it can also increase vulnerability to extremely low generation at the country level.

Our results support the recommendation to bias-correct reanalysis wind speeds \cite{gruber_towards_2022}, provided that the correction is regionally validated. Gaussian smoothing of the power curve calibrated against historical data \cite{andresen_validation_2015,staffell_using_2016} absorbs several effects in one set of parameters (Sec.~\ref{sec:supp_gaussian_smoothing}). Evaluating each assumption individually instead shows how far bias-corrected wind speeds alone can improve the agreement with historical duration curves (Fig.~\ref{fig:cf_distribution_with_reference}).

\subsection{Implications for energy system modellers}
Modelling choices should be evaluated against the low-generation tail, not only against the mean. As their relative importance differs between mean generation, VRE droughts, and system outcomes, we group the following recommendations by effect size, starting with those that matter across all metrics.

\textit{Use multiple weather years.} Inter-annual variability is comparable to the most influential modelling choices and exceeds most others examined here. Single-year assessments or representative days risk underestimating uncertainty and missing system-relevant stress periods.

\textit{Bias-correct the input (and validate it regionally).} Correcting wind speeds matters more than correcting solar irradiance, and for wind, the choice of method matters little. For solar, quantile mapping lowers winter output and raises backup requirements. As the quality of a wind speed correction varies between regions (Fig.~\ref{fig:cf_distribution_with_reference}), a single correction dataset applied without regional validation may perform worse than no correction.

\textit{Represent resource heterogeneity.} Resource quality matters considerably more than the capacity layout within a country, especially for onshore wind. Where high spatial resolution \cite{frysztacki_strong_2021,jacobson_quantifying_2024} is computationally infeasible, several resource grades per country are a practical alternative; in our system analysis, three grades already reduce total installed capacity by \qtyrange{9}{12}{\percent} compared with a single country-level time series.

\textit{Include forward-looking technologies.} Higher hub heights are especially valuable during droughts (Fig.~\ref{fig:drought_delivered_onwind_seasons_europe_avg}), whereas larger turbines have a smaller impact. HSAT affects mean yield and drought behaviour in opposite directions (Figs.~\ref{fig:solar-cf-yearly} and~\ref{fig:drought_delivered_solar_seasons_europe_avg}), so it must be represented explicitly.

\textit{Include wake losses (even in simplified form).} Their effect is modest onshore, but offshore, where plants are more densely spaced, they reduce mean capacity factors by \qtyrange{8}{13}{\percent} and deepen shortfalls during VRE droughts (Figs.~\ref{fig:offshore-cf-yearly} and~\ref{fig:drought_delivered_offwind-ac_seasons_europe_avg}). At the system level, including them increases total installed capacity by \qtyrange{6}{8}{\percent}. As we consider only losses within a single representative wind farm, these effects are likely conservative: with the dense offshore clusters and energy hubs planned towards 2050, large-scale wakes between neighbouring wind farms are expected to increase losses considerably \cite{bramstoft_role_2026}.

\textit{Choose wind interpolation methods pragmatically.} As long as \qty{100}{\meter} wind speeds are available, the differences between logarithmic and power-law extrapolation and model-level winds are small at the aggregated country level.

\subsection{Limitations and outlook}
\label{sec:limitations}
A key limitation of this study is the one-at-a-time sensitivity design. Although it is transparent and allows modelling choices to be ranked by their influence, it cannot capture interactions among them. The system analysis is conducted at the country level, neglecting intra-national transmission constraints and local balancing effects, which influence system performance \cite{frysztacki_strong_2021}. It also assumes perfect foresight and thus ignores forecast errors, which may further increase flexibility requirements. As an electricity-only model, it omits electrified heating, transport, and industrial demand and thus compound events of low renewable generation and high heating demand \cite{gotske_designing_2024}. The system stress reported here is therefore likely underestimated. VRE drought characteristics also depend on the detection methodology \cite{kittel_measuring_2024,biewald_evaluation_2025}; our conclusions are robust to thresholds between the 5th and 15th percentiles (Sec.~\ref{sec:supp_threshold}), whereas minimum-duration and merging rules were kept fixed.

We compare installed capacities rather than total system costs. Future costs of higher hub heights and larger turbines are uncertain, and additional cost assumptions would make it harder to isolate the effects of the resource modelling choices; we therefore keep investment costs constant across configurations, except for HSAT, for which specific cost assumptions are available. The system-level benefits of the hub-height and turbine configurations reflect technical performance rather than economic competitiveness and should be interpreted as upper bounds.

Capacity layouts, turbine types, and PV orientations are deliberately generic rather than matched to the historical fleet, which likely explains part of the deviation from reported CFs. Validation is further limited by the historical data: installed capacities and generation differ across sources and are affected by unreported curtailment, and solar CFs refer to module (DC) capacity, even though systems are frequently oversized relative to their inverters. A consistent, open dataset of hourly generation, curtailment, and installed capacity by technology would substantially strengthen such validation.

These limitations point to three main directions for future work. First, a global sensitivity analysis could evaluate interactions among modelling choices. Second, the system study could be extended to a higher spatial resolution and to sector coupling to investigate whether these amplify or dampen the most critical periods. Third, climate change projections, which are outside the scope of this work, could show whether future shifts in renewable resources exceed both inter-annual variability and the effects of the modelling choices examined here. This is particularly relevant because infrastructure decisions made over the next decade will shape the European energy system for much longer.

\subsection{Concluding remarks}
Renewable modelling choices are rarely examined systematically in adequacy assessments and energy system studies. However, they alter the electricity delivered during VRE droughts and the resulting system design, in some cases by amounts comparable to inter-annual weather variability. As multi-decadal weather records become standard in adequacy assessments such as ERAA \cite{entso-e_european_2025}, the conversion of these data into generation time series deserves the same scrutiny as the choice of weather data itself. The same principle extends beyond modelling: siting decisions and support schemes based solely on annual generation may undervalue resources that perform well during periods of scarcity. A resilient European power system, therefore, requires renewable resources to be assessed by what they deliver during the scarcest periods, not only by what they deliver on average.

%% file: chapters/05_supplementary.tex
\section*{Supplementary Material}
\pdfbookmark[1]{Supplementary Material}{supplementary}

\subsection*{Abbreviations}
{\small
\begin{longtable}{@{}p{0.2\linewidth}p{0.76\linewidth}@{}}
\toprule
Abbreviation & Meaning \\
\midrule
\endhead
\bottomrule
\endfoot

ATB     & Annual Technology Baseline (NREL) \\
CDF     & Cumulative distribution function \\
CF      & Capacity factor \\
CMIP6   & Coupled Model Intercomparison Project, Phase 6 \\
CORINE  & Coordination of Information on the Environment (land cover dataset) \\
DE, DK, ES, IT & Germany, Denmark, Spain, Italy (ISO 3166 country codes) \\
ECMWF   & European Centre for Medium-Range Weather Forecasts \\
ENTSO-E & European Network of Transmission System Operators for Electricity \\
ERA5    & Fifth-generation ECMWF atmospheric reanalysis \\
ERAA    & European Resource Adequacy Assessment \\
GEBCO   & General Bathymetric Chart of the Oceans \\
GWA, GWA v2, GWA v4 & Global Wind Atlas (versions 2 and 4) \\
H\textsubscript{2} & Hydrogen \\
HSAT    & Horizontal single-axis tracking \\
IRENA   & International Renewable Energy Agency \\
NREL    & National Renewable Energy Laboratory \\
p5      & 5th percentile \\
PHS     & Pumped hydro storage \\
pp      & Percentage points \\
PV      & Photovoltaic(s) \\
SARAH-3 & Surface Solar Radiation Data Set -- Heliosat, edition 3 \\
VRE     & Variable renewable energy \\
\end{longtable}

\subsection*{Symbols}
\begin{longtable}{@{}p{0.2\linewidth}p{0.6\linewidth}p{0.14\linewidth}@{}}
\toprule
Symbol & Description & Unit \\
\midrule
\endhead
\bottomrule
\endfoot
\multicolumn{3}{@{}l}{\emph{Indices}} \\
$i$ & Grid cell & -- \\
$t$ & Hour (time step) & -- \\
$c$ & Country & -- \\
$d$ & Day of the year & -- \\
$m$ & Calendar month & -- \\
$e$ & Drought event & -- \\
$j$, $J$ & GWA pixel within an ERA5 cell, number of such pixels & -- \\
$\ell$ & ERA5 model level & -- \\
\addlinespace
\multicolumn{3}{@{}l}{\emph{Bias correction}} \\
$x^{\text{ERA5}}_{i,t}$, $x^{\text{corr}}_{i,t}$ & ERA5 and bias-corrected value (wind speed or irradiance) & varies \\
$\overline{x}^{\text{ERA5}}_i$, $\overline{x}^{\text{ref}}_i$ & Mean of ERA5 and of the reference dataset over the bias-correction period & varies \\
$\beta_i$ & Linear correction factor & -- \\
$F^{\text{ERA5}}_{m}$ & CDF of the ERA5 data in month $m$ & -- \\
$Q^{\text{ref}}_{m}$ & Quantile function (inverse CDF) of the reference data in month $m$ & varies \\
$F(U;A,k)$ & Weibull CDF of wind speed & -- \\
$A$, $k$ & Weibull scale and shape parameters & \unit{\meter\per\second}, -- \\
$(A_j, k_j)$ & Sector-averaged Weibull parameters of GWA pixel $j$ & \unit{\meter\per\second}, -- \\
$F^{\text{mix,ref}}$ & Average CDF of the GWA pixels in an ERA5 cell & -- \\
$F^{\text{ref}}$ & Representative reference Weibull distribution & -- \\
$\hat{A}_{\text{ERA5}}$, $\hat{k}_{\text{ERA5}}$, $\hat{A}_{\text{ref}}$, $\hat{k}_{\text{ref}}$ & Fitted Weibull parameters of ERA5 and of the reference & \unit{\meter\per\second}, -- \\
$\tau_{i,t}$ & Cumulative probability of $U_{i,t}$ under the ERA5 distribution & -- \\
$I_{\text{SARAH}}$, $I_{\text{ERA5}}$ & Irradiance from SARAH-3 and ERA5 & \unit{\watt\per\square\meter} \\
\addlinespace
\multicolumn{3}{@{}l}{\emph{Wind speed at hub height}} \\
$u_{i,t}$, $v_{i,t}$ & Zonal and meridional wind components & \unit{\meter\per\second} \\
$U_{i,t}$ & Wind speed & \unit{\meter\per\second} \\
$U_{10,i,t}$, $U_{100,i,t}$ & Wind speed at \qty{10}{\meter} and \qty{100}{\meter} & \unit{\meter\per\second} \\
$U^{\text{corr}}_{i,t}$ & Bias-corrected wind speed & \unit{\meter\per\second} \\
$U_{\text{hub},i,t}^{\log}$, $U_{\text{hub},i,t}^{\text{pow}}$, $U_{\text{hub},i,t}^{\text{ml}}$ & Hub-height wind speed from the logarithmic law, power law, and model-level interpolation & \unit{\meter\per\second} \\
$h_{\text{hub}}$ & Hub height & \unit{\meter} \\
$z_{0,i,t}$ & Surface roughness length & \unit{\meter} \\
$\alpha_{i,t}$ & Wind shear exponent & -- \\
$z_{\ell,i,t}$, $z^{\text{top}}_{\ell,i,t}$ & Height above ground of model level $\ell$ and of the top of its layer & \unit{\meter} \\
$\Delta z_{\ell,i,t}$ & Thickness of model layer $\ell$ & \unit{\meter} \\
$p_{\text{s},i,t}$ & Surface pressure & \unit{\pascal} \\
$p^{\text{bot}}_{\ell,i,t}$, $p^{\text{top}}_{\ell,i,t}$ & Pressure at the bottom and top of model layer $\ell$ & \unit{\pascal} \\
$T_{\ell,i,t}$, $T_{\text{v},\ell,i,t}$ & Temperature and virtual temperature & \unit{\kelvin} \\
$q_{\ell,i,t}$ & Specific humidity & \unit{\kilo\gram\per\kilo\gram} \\
$R_{\text{d}}$ & Specific gas constant of dry air & \unit{\joule\per\kilo\gram\per\kelvin} \\
$g$ & Gravitational acceleration & \unit{\meter\per\second\squared} \\
$\varepsilon$ & Ratio of the gas constants of dry air and water vapour ($=0.622$) & -- \\
\addlinespace
\multicolumn{3}{@{}l}{\emph{Spatial representation}} \\
$\overline{CF}_i$ & Mean capacity factor of cell $i$ over 1941--2025 & -- \\
$w_i$ & Layout weight of cell $i$ & \unit{\mega\watt} \\
$\gamma$ & Layout exponent & -- \\
$S_i$ & Eligible area of cell $i$ & \unit{\square\kilo\meter} \\
$\kappa_i^{\max}$ & Maximum installable capacity per unit area & \unit{\mega\watt\per\square\kilo\meter} \\
$\overline{CF}^{\min}_c$, $\overline{CF}^{\max}_c$ & Lowest and highest cell-mean capacity factor in country $c$ & -- \\
$R_c$ & Range of cell-mean capacity factors in country $c$ & -- \\
\addlinespace
\multicolumn{3}{@{}l}{\emph{VRE droughts}} \\
${CF}^{\text{thr}}_{c,d}$ & Detection threshold (smoothed day-of-year 5th percentile) & -- \\
${CF}^{\text{norm}}_{c,d}$ & Seasonal norm (smoothed day-of-year reference capacity factor) & -- \\
$\overline{CF}_e$ & Mean capacity factor during event $e$ & -- \\
$\overline{CF}^{\text{norm}}_e$ & Mean seasonal norm during event $e$ & -- \\
$L_e$ & Duration of event $e$ & \unit{\hour} \\
$\delta_e$ & Delivered share of electricity during event $e$ & \unit{\percent} \\
$\bar{\delta}_c$ & Aggregate (length-weighted) delivered share per country, technology, and season & \unit{\percent} \\
\end{longtable}
}
\clearpage

\subsection{Literature gap}
Figure~\ref{fig:gap_matrix} places this study in the context of previous modelling studies that compare alternative renewable modelling choices within a single framework. Existing studies vary only subsets of these choices, and none evaluates them jointly across capacity factors, VRE droughts, and system outcomes.

\begin{figure}[htbp]
    \centering
    \renewcommand{\sfdefault}{phv}% Helvetica, as in the other figures
    \input{figures/method/figure1_gap_matrix.pgf}
    \caption{Renewable modelling choices examined in previous modelling studies compared with this study. Circles mark modelling choices for wind (blue) and solar PV (yellow); squares mark the metrics on which their effect is evaluated: capacity factors or generation (CF/gen.), VRE droughts, and power system outcomes (System). Filled symbols indicate that a study varies or compares the choice; grey symbols indicate choices that are held fixed or not considered. The $^{\star}$ marks global or non-European studies. The selection is illustrative rather than exhaustive and comprises modelling studies that compare alternative choices within a single framework.}
    \label{fig:gap_matrix}
\end{figure}

\subsection{Baseline configuration}
\label{sec:baseline_config}

The baseline uses ERA5 without bias correction and assumes static, south-oriented, \qty{35}{\degree}-tilt silicon solar PV modules, a \qty{3}{\mega\watt} onshore wind turbine with an \qty{80}{\meter} hub height (Vestas V112, \qty{305}{\watt\per\square\meter} specific power), and a \qty{5.5}{\mega\watt} offshore turbine with a \qty{120}{\meter} hub height (NREL 2020 ATB Reference, \qty{229}{\watt\per\square\meter}). Wind speed is extrapolated from \qty{100}{m} to hub height with the logarithmic law (Sec.~\ref{sec:supp_wind_height}), capacity is distributed with the proportional layout ($\gamma=1$; Sec.~\ref{sec:spatial}), and wake losses are not considered. 

Eligible land follows the exclusions listed in Fig.~\ref{fig:methods_overview}. For each technology, protected areas (Natura 2000 \cite{european_environment_agency_eea_natura_2022}) are excluded, and eligible land is determined from CORINE land cover \cite{european_environment_agency_eea_corine_2022}. Offshore wind can only be installed at locations with a water depth of less than \qty{60}{\meter} based on the bathymetry dataset GEBCO \cite{weatherall_new_2015}. Furthermore, the maximum installation density $\kappa_i^{\max}$ is restricted to \qty{5.1}{\mega\watt\per\square\kilo\meter} for utility-scale solar PV (\qty{4.43}{\mega\watt\per\square\kilo\meter} with HSAT; Sec.~\ref{sec:supp_tech_wake}), \qty{3}{\mega\watt\per\square\kilo\meter} for onshore, and \qty{2}{\mega\watt\per\square\kilo\meter} for offshore wind. These limits represent average capacities across all eligible areas and should therefore not be interpreted as solar- or wind-farm densities, which are substantially higher within individual projects. An extensive description of exclusion zones and available areas is provided in Supplementary Material S6 from Neumann et al.~\cite{neumann_potential_2023}.

\subsection{Bias correction}
\label{sec:supp_bias}

\subsubsection{Reference datasets}
For solar irradiance, we use the SARAH-3 satellite dataset as the reference \cite{pfeifroth_surface_2023}. We calculate the bias by comparing the period 2011--2020 in SARAH-3 with the same period in ERA5. The native \qty{0.05}{\degree} SARAH-3 data are re-gridded to the ERA5 resolution using area-weighted average resampling. Direct and diffuse irradiance are corrected separately.

\begin{figure}[htbp]
  \centering
  \includegraphics[width=0.89\linewidth]{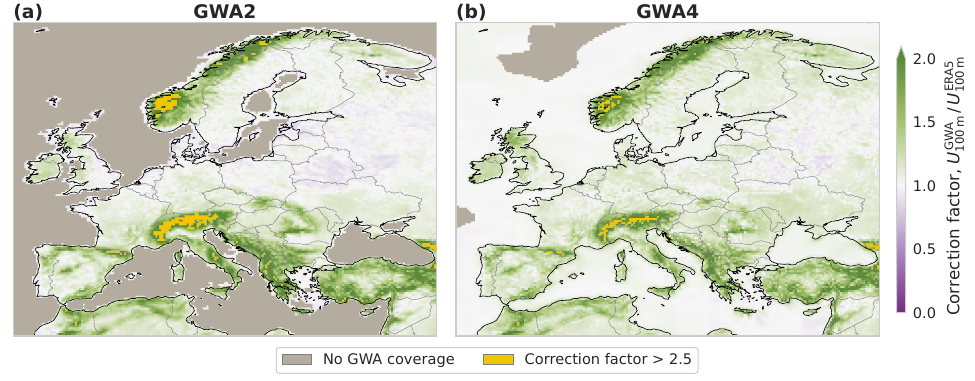}
  \caption{Spatial distribution of linear correction factors for wind speed at a height of \qty{100}{\meter}, derived from GWA v2 and GWA v4, compared with ERA5. Values greater than one indicate upward corrections to the ERA5 wind speeds. The coverage in panel (a) (GWA v2) is limited to land and near-coastal offshore areas, while panel (b) (GWA v4) covers a much larger area. Both show substantial regional variations, with the strongest upward corrections occurring in areas with complex terrain. Grey indicates regions without GWA coverage.}
  \label{fig:bias_factor_gwa2_gwa4_era5}
\end{figure}

\begin{figure}[htbp]
  \centering
  \includegraphics[width=0.6\linewidth]{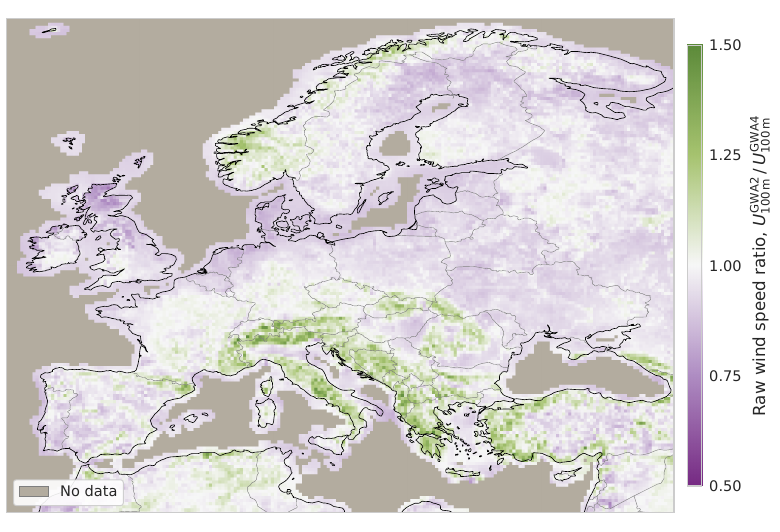}
  \caption{Spatial distribution of linear wind-speed bias correction factors at \qty{100}{\meter} derived from GWA v4 and GWA v2.
  A factor below 1 means GWA v4 has, on average, a higher reported wind speed than GWA v2. The correction factors exhibit substantial regional variability. Complex mountainous terrain has higher wind speeds in GWA v2. By contrast, flat areas and coastal regions show, on average, lower wind speeds in GWA v2 than in GWA v4. This explains why bias-correcting ERA5 wind speeds with GWA v2 instead of v4 yields CFs closer to the reported values in Denmark and Germany. Grey areas denote regions without GWA v2 coverage.}
  \label{fig:bias_factor_gwa2_gwa4}
\end{figure}

For wind speed, we calculate the bias by comparing the period 2008--2017 in the microscale data from GWA v4 \cite{davis_global_2023} with the corresponding years in ERA5. The native \qty{250}{\meter} resolution data are re-gridded to the ERA5 resolution (\qty{0.25}{\degree} $\times$ \qty{0.25}{\degree}, corresponding to a cell size of $\approx$~\qty{30}{\kilo\meter} $\times$ \qty{30}{\kilo\meter}) via area-weighted average resampling. We correct only the \qty{100}{m} wind speed. The \qty{10}{m} wind speed is left uncorrected, because ERA5 does not predict it directly but derives it from the lowest model level with an exposure correction intended to approximate station observations \cite{betts_near-surface_2019}, whereas GWA near-surface winds reflect local roughness and orography \cite{davis_global_2023}. Surface roughness (introduced in Sec.~\ref{sec:wind_height}) is not bias-corrected either. Because the evidence for improved accuracy refers to GWA v2 \cite{murcia_validation_2022,nayak_validation_2025}, Figs.~\ref{fig:bias_factor_gwa2_gwa4_era5} and~\ref{fig:bias_factor_gwa2_gwa4} compare the correction factors derived from both versions.

\subsubsection{Linear correction}

\begin{figure}[htbp]
  \centering
  \includegraphics[width=0.89\linewidth]{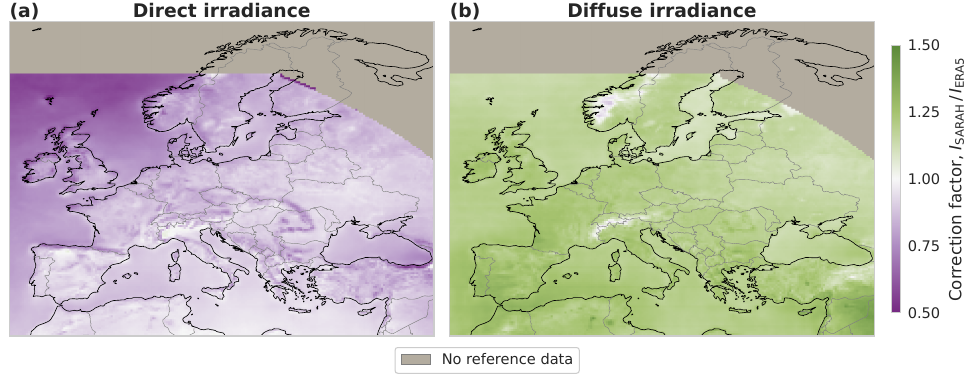}
  \caption{Maps of linear correction factors for (a) direct irradiance and (b) diffuse irradiance across Europe. The correction factor is calculated as the ratio $I_{\mathrm{SARAH}}/I_{\mathrm{ERA5}}$. Purple shades indicate correction factors below 1, green shades indicate correction factors above 1, and white corresponds to values close to unity. Grey areas mark regions without reference data, which are excluded from the bias correction.}
  \label{fig:solar_bias_correction_map}
\end{figure}

The ERA5 time series $x^{\text{ERA5}}_{i,t}$ for grid cell $i$ and hour $t$ is rescaled by a correction factor $\beta_i$, defined as the ratio of the reference mean $\overline{x}^{\text{ref}}_i$ to the ERA5 mean $\overline{x}^{\text{ERA5}}_i$ over the bias-correction period:
\begin{equation}
x^{\text{corr}}_{i,t} = \beta_i\, x^{\text{ERA5}}_{i,t}, \qquad
\beta_i = \frac{\overline{x}^{\text{ref}}_i}{\overline{x}^{\text{ERA5}}_i}.
\end{equation}
The factor $\beta_i$ is determined independently for each grid cell and applied uniformly to all time steps and weather years. Maps of $\beta_i$ are shown in Fig.~\ref{fig:bias_correction_era5_gwa4_vertical}a for the \qty{100}{m} wind speed and in Fig.~\ref{fig:solar_bias_correction_map} for direct and diffuse irradiance. Over the Alps and the Norwegian mountains, the solar correction factors remain close to unity, highlighting the influence of cloud cover and its better representation in satellite data than in physical reanalysis data.

\subsubsection{Quantile mapping}
\label{sec:qm_correction}
Quantile mapping adjusts the value at each percentile to match the reference distribution, rather than applying a uniform scaling to the entire distribution. The shape of the corrected distribution, including both tails, therefore follows the reference dataset \cite{cannon_bias_2015}. In this study, the quantile mapping approach differs between wind and solar, reflecting the stronger seasonal dependence of solar irradiance and the limited availability of wind data for correction.

\begin{figure}[htbp]
    \centering
    \includegraphics[width=0.80\linewidth]{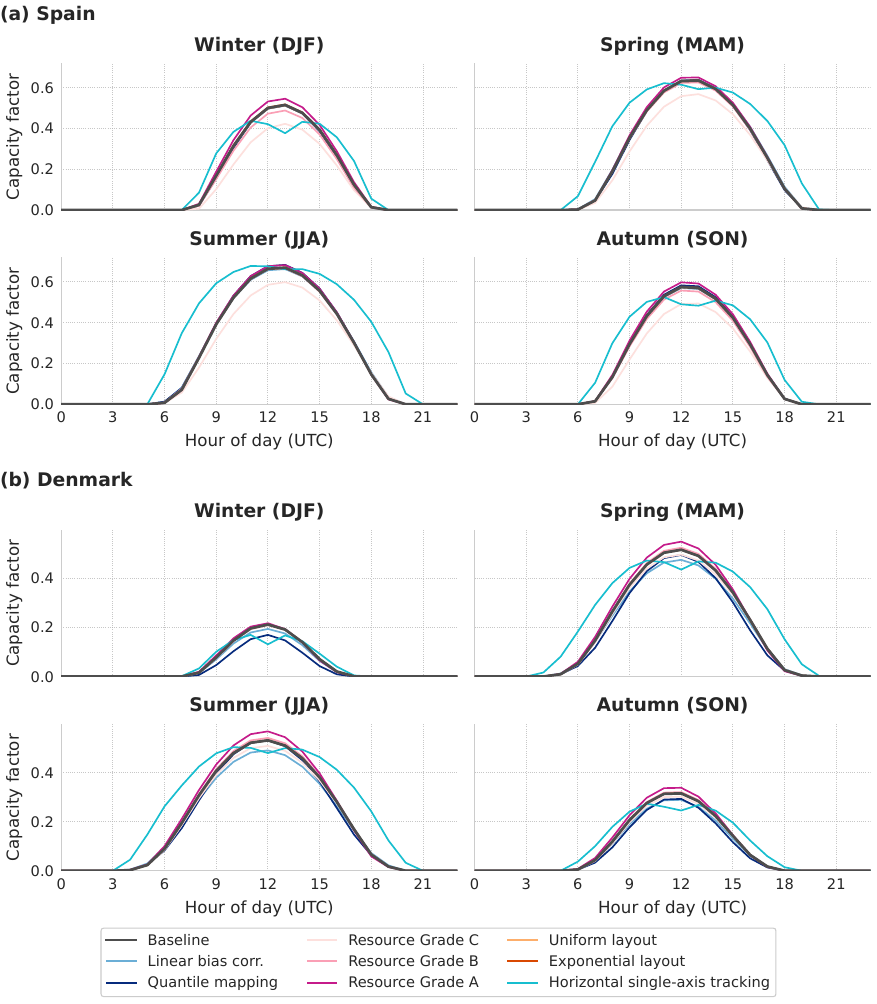}
    \caption{Seasonal mean diurnal cycles of solar capacity factor for (a)~Spain and (b)~Denmark under different modelling configurations. Bias correction (linear and quantile mapping), capacity layout, and resource grades have only a minor effect on the diurnal shape in Spain, while in Denmark the corrections are more pronounced, particularly in winter, when quantile mapping substantially lowers midday capacity factors. Horizontal single-axis tracking systematically broadens the profile, shifting generation into morning and evening hours at the cost of a reduced midday peak in low-sun seasons.}
    \label{fig:diurnal_cycle_solar}
\end{figure}

\begin{figure}[htbp]
  \centering
  \includegraphics[width=0.89\linewidth]{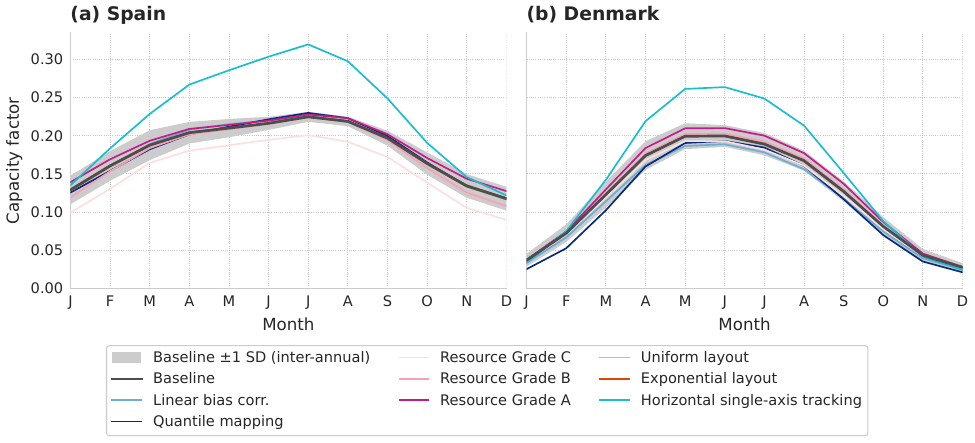}
  \caption{Monthly mean solar capacity factor for (a)~Spain and (b)~Denmark under all modelling configurations, averaged over the 85 weather years. The shaded band shows the baseline's inter-annual variability ($\pm$1 standard deviation of the monthly means). HSAT raises the summer capacity factor the most, while bias correction and capacity layout change the level only marginally and leave the seasonal shape unchanged.}
  \label{fig:seasonal_cycle_solar}
\end{figure}

For solar, monthly distributions of hourly irradiance are calculated for the bias-correction period (2011--2020) using re-gridded SARAH-3 data and ERA5. The ERA5 time series $x^{\text{ERA5}}_{i,t}$ is then bias-corrected per month by mapping each ERA5 value to the corresponding quantile of the reference distribution:

\begin{equation}
x^{\text{corr}}_{i,t} = Q^{\text{ref}}_{m}\!\left(F^{\text{ERA5}}_{m}(x^{\text{ERA5}}_{i,t})\right),
\end{equation}

where $F^{\text{ERA5}}_{m}$ denotes the cumulative distribution function (CDF) of ERA5 data for each month $m$, and $Q^{\text{ref}}_{m}$ is the corresponding quantile function (the inverse CDF) of the reference dataset (SARAH-3). The correction is computed independently for each grid cell and for each variable (direct and diffuse irradiance). In practice, the adjustment is implemented using \texttt{xsdba} v0.5.0 \cite{dupuis_xsdba_2025}. Quantile mapping does not distort the solar diurnal cycle (Fig.~\ref{fig:diurnal_cycle_solar}) or the seasonal profile (Fig.~\ref{fig:seasonal_cycle_solar}).

For wind, GWA v4 provides only static Weibull parameters, so quantile mapping cannot be applied directly. Instead, a reference CDF is constructed for each grid cell from these parameters. The Weibull CDF is

\begin{equation}
    F(U;\,A,k) = 1 - \exp\!\left[-\left(\frac{U}{A}\right)^{k}\right], \quad U \ge 0,
\end{equation}
where $U$ denotes wind speed (\unit{\meter\per\second}) and $A>0$ and $k>0$ are the scale and shape parameters. For each grid cell, we estimate the ERA5 Weibull parameters $(\hat{A}_{\text{ERA5}}, \hat{k}_{\text{ERA5}})$ by maximum likelihood from the 10-year hourly \qty{100}{m} wind speed time series of the bias-correction period. For the reference distribution, GWA provides sector-averaged Weibull parameters $(A_j, k_j)$ for each of the $J$ \qty{250}{\meter} pixels within an ERA5 cell. We combine them into a single reference CDF by averaging the pixel CDFs:
\begin{equation}
    F^{\text{mix,ref}}(U) = \frac{1}{J}\sum_{j=1}^{J}
    \left(1 - \exp\!\left[-\left(\frac{U}{A_j}\right)^{k_j}\right]\right).
\end{equation}

\begin{figure}[htbp]
  \centering
  \includegraphics[width=0.80\linewidth]{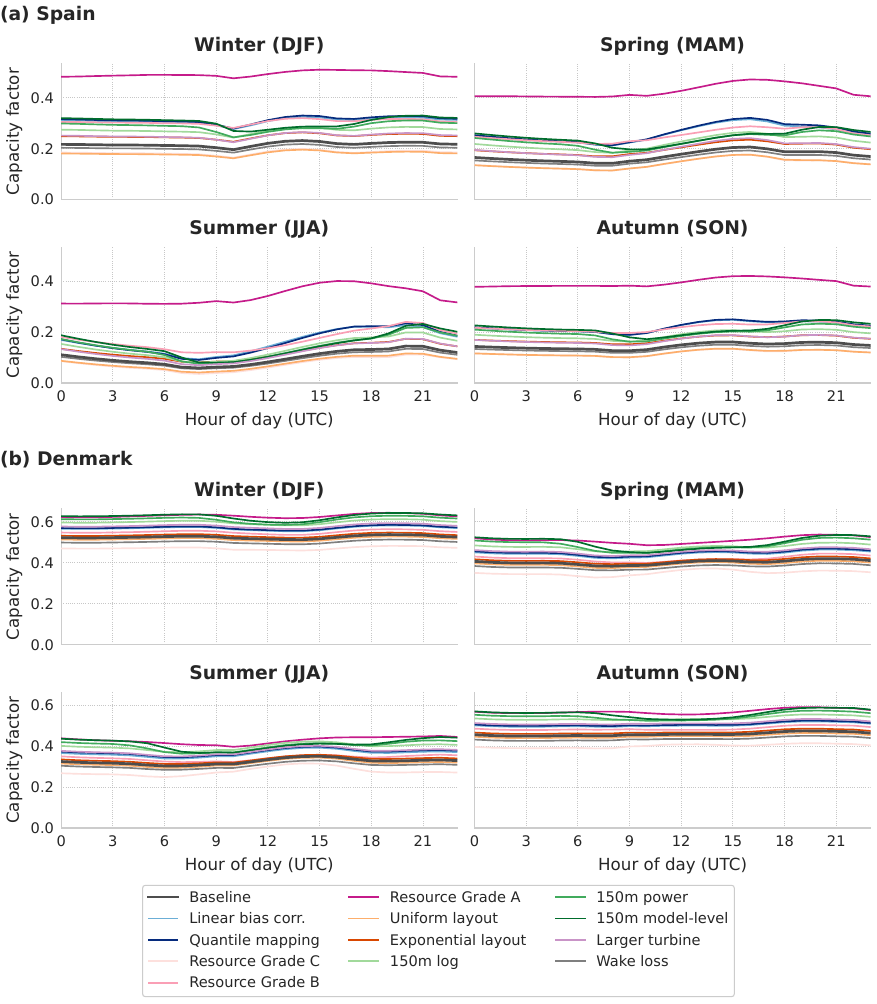}
  \caption{Seasonal mean diurnal cycles of onshore wind capacity factor for (a)~Spain and (b)~Denmark under all modelling configurations. The configurations shift the level of the diurnal cycle but hardly change its shape.}
  \label{fig:diurnal_cycle_onwind}
\end{figure}

\begin{figure}[htbp]
  \centering
  \includegraphics[width=0.89\linewidth]{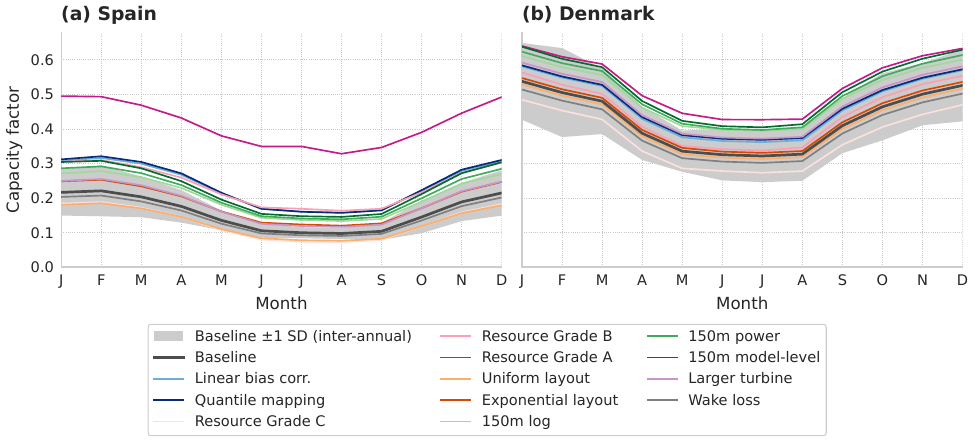}
  \caption{Monthly mean onshore wind capacity factor for (a)~Spain and (b)~Denmark under all modelling configurations, averaged over the 85 weather years; shaded band as in Fig.~\ref{fig:seasonal_cycle_solar}. All configurations preserve the seasonal cycle with a winter maximum and a summer minimum, and mainly shift its level, most strongly for Resource Grade~A in Spain.}
  \label{fig:seasonal_cycle_onwind}
\end{figure}

This average is not itself a Weibull distribution. A representative Weibull $F^{\text{ref}}$ with parameters $(\hat{A}_{\text{ref}}, \hat{k}_{\text{ref}})$ is therefore fitted to $F^{\text{mix,ref}}$ using nonlinear least squares over 200 quantile points in $[0.01,\,0.99]$, which restores the closed form used below.
Bias correction is performed by mapping the \qty{100}{m} ERA5 wind speed $U_{i,t}$ to its cumulative probability $\tau_{i,t}$ under the ERA5 distribution and evaluating the reference quantile function at this probability:

\begin{equation}
\tau_{i,t} = F^{\text{ERA5}}\!\left(U_{i,t};\,\hat{A}_{\text{ERA5}},\hat{k}_{\text{ERA5}}\right), \qquad
    U^{\text{corr}}_{i,t} = Q^{\text{ref}}\!\left(\tau_{i,t};\,\hat{A}_{\text{ref}},\hat{k}_{\text{ref}}\right).
\end{equation}
Using $-\ln(1 - F(U)) = (U/A)^k$, this yields the closed-form expression
\begin{equation}
    U^{\text{corr}}_{i,t} = \hat{A}_{\text{ref}}
    \left(\frac{U_{i,t}}{\hat{A}_{\text{ERA5}}}\right)^{\hat{k}_{\text{ERA5}}/\hat{k}_{\text{ref}}}.
\end{equation}
The wind at an ERA5 grid cell is corrected only if all three of the following conditions are met: (i) more than \qty{80}{\percent} of its GWA pixels are valid, (ii) the Weibull fits converge, and (iii) $\hat{k}_{\text{ref}} \ge 1.25$. Otherwise, the original ERA5 values are retained. Diurnal and seasonal profiles of the resulting onshore wind capacity factors (Figs.~\ref{fig:diurnal_cycle_onwind} and~\ref{fig:seasonal_cycle_onwind}) confirm that quantile mapping does not unintentionally modify the temporal patterns. Correction factors and transfer functions are available on Zenodo~\cite{karkossa_bias-correction_2026}.

\subsection{Estimation of wind speed at hub height}
\label{sec:supp_wind_height}

In all three methods below, the wind speed $U_{i,t}$ at a given height is first obtained as the magnitude of the horizontal wind vector, from its eastward $u_{i,t}$ and northward $v_{i,t}$ components,

\begin{equation}
U_{i,t} = \sqrt{u_{i,t}^2 + v_{i,t}^2}.
\label{eq:wspd}
\end{equation}

Method (i) uses the ERA5 wind speed at \qty{100}{m}, $U_{100,i,t}$; method (ii) additionally uses the wind speed at \qty{10}{m}, $U_{10,i,t}$. Both are obtained from Eq.~\eqref{eq:wspd}. The quantity to be determined is the wind speed $U_{\text{hub},i,t}$ at hub height $h_{\text{hub}}$. 

\paragraph{(i) Logarithmic law extrapolation.}
Assuming neutral atmospheric stratification, we use the forecast surface roughness field from ERA5, which provides an hourly roughness length $z_{0,i,t}$ for each grid cell. At each grid cell and time, we take the \qty{100}{m} wind speed $U_{100,i,t}$ and apply the logarithmic law to extrapolate it to hub height $U_{\text{hub},i,t}$:

\begin{equation}
U_{\text{hub},i,t}^{\log}
  = U_{100,i,t}\,
    \frac{\ln\!\big(h_{\text{hub}} / z_{0,i,t}\big)}
         {\ln\!\big(\qty{100}{\meter} / z_{0,i,t}\big)} .
\end{equation}
This approach, adapted from Andresen et al.~\cite{andresen_validation_2015}, is used in the baseline.

\paragraph{(ii) Power-law extrapolation.}
From the ERA5 \qty{10}{m} and \qty{100}{m} wind speeds, we estimate an hourly shear exponent

\begin{equation}
\alpha_{i,t} = \frac{\ln\!\big(U_{100,i,t}/U_{10,i,t}\big)}{\ln(\qty{100}{\meter}/\qty{10}{\meter})},
\end{equation}

and compute

\begin{equation}
U_{\text{hub},i,t}^{\text{pow}}
  = U_{100,i,t}\left(\frac{h_{\text{hub}}}{\qty{100}{\meter}}\right)^{\alpha_{i,t}}.
\end{equation}

\paragraph{(iii) Model-level interpolation.}
Wind speed at hub height, $U_{\text{hub}, i,t}^{\text{ml}}$, is calculated by linearly interpolating between the two ERA5 model levels directly above and below the hub height. Model levels follow the terrain and are defined by pressure rather than height, so their heights above ground, $z_{\ell, i,t}$, vary in space and time and must be reconstructed for each grid cell $i$ and time step $t$. For this, hourly temperature $T_{\ell, i,t}$ and specific humidity $q_{\ell, i,t}$ are retrieved alongside the wind components $u_{\ell, i,t}$ and $v_{\ell, i,t}$ at model levels $\ell = 122,\dots,137$, on the same \qty{0.25}{\degree} $\times$ \qty{0.25}{\degree} grid as the single-level fields, together with the surface pressure $p_{\text{s}, i,t}$.
Each model layer is bounded by two half levels, each with pressure $a + b\,p_{\text{s}, i,t}$, where $a$ and $b$ are the tabulated ECMWF hybrid coefficients of that half level \cite{ecmwf_l137_2026}. With $p^{\text{top}}_{\ell,i,t}$ and $p^{\text{bot}}_{\ell,i,t}$ the pressures at the upper and lower boundaries of layer $\ell$, the layer thickness $\Delta z_{\ell, i,t}$ follows from the hypsometric equation together with the virtual temperature $T_{\text{v},\ell,i,t}$,
\begin{equation}
\Delta z_{\ell,i,t} = \frac{R_{\text{d}}}{g}\, T_{\text{v},\ell,i,t}\,
  \ln\!\left(\frac{p^{\text{bot}}_{\ell,i,t}}{p^{\text{top}}_{\ell,i,t}}\right),
\qquad
T_{\text{v},\ell,i,t} = T_{\ell,i,t}\,
  \frac{q_{\ell,i,t}+\varepsilon}{\varepsilon\,(1+q_{\ell,i,t})} ,
\end{equation}
with $R_{\text{d}}$ the specific gas constant for dry air, $g$ the gravitational acceleration, $T_{\ell,i,t}$ the temperature, $q_{\ell,i,t}$ the specific humidity, and $\varepsilon = 0.622$. Summing these thicknesses upward from the surface gives the height above ground of the top of
each layer,
\begin{equation}
    z^{\text{top}}_{\ell,i,t} = \sum_{k=\ell}^{137} \Delta z_{k,i,t},
\end{equation}
and the height $z_{\ell,i,t}$ of model level $\ell$ is taken as the midpoint between
the top and bottom of its layer.

The wind speed $U_{\ell,i,t}$ on each model level then follows from Eq.~\eqref{eq:wspd} and is interpolated linearly in height to the hub height,
\begin{equation}
U_{\text{hub},i,t}^{\text{ml}}
  = \operatorname{interp}_z\!\Big(\,U_{\ell,i,t}\,;\ z_{\ell,i,t} \rightarrow h_{\text{hub}}\Big).
\end{equation}

This follows the approach of Hahmann et al.~\cite{hahmann_current_2022}, using linear interpolation in height and applying it to ERA5 rather than CMIP6 output.

These last two operations are not interchangeable. Interpolating the two wind components first and taking the magnitude afterwards yields, by the triangle inequality, a speed no larger than the one obtained here, since the vectors partially cancel wherever the wind direction veers between the two levels. Taking the magnitude on each model level first avoids this cancellation. Given the close vertical spacing of the ERA5 model levels at these heights, the resulting difference remains negligible.

\subsection{Resource grades}
\label{sec:supp_spatial}
 
\begin{figure}[htbp]
  \centering
  \includegraphics[width=0.89\linewidth]{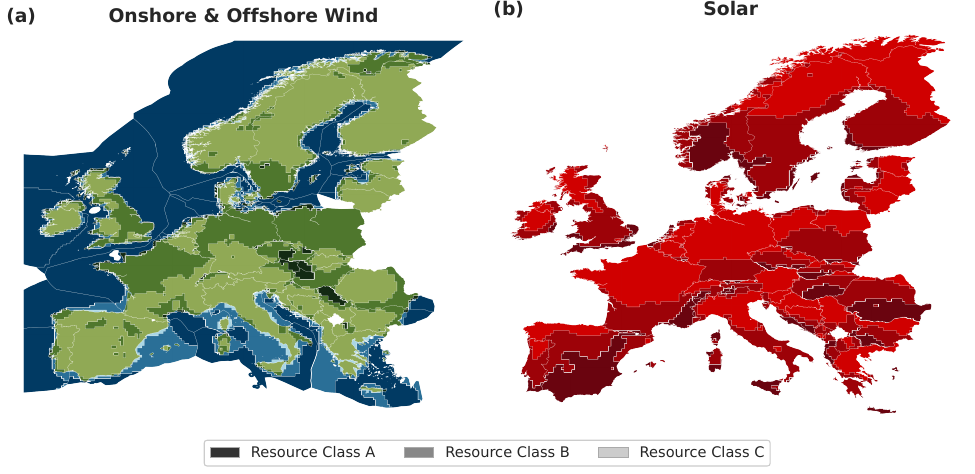}
  \caption{Spatial distribution of renewable resource grades across Europe. (a)~Onshore (green) and offshore (blue) wind and (b)~solar resource quality per country, classified into three grades from Grade~A (best, darkest) to Grade~C (worst, lightest).}
  \label{fig:resource_grade_classification}
\end{figure}

To retain part of the resource spread within a country, we divide the range between each country's poorest and best cells, $\overline{CF}^{\min}_c$ and $\overline{CF}^{\max}_c$, into three equal intervals: cells in the upper third form Grade~A, the middle third form Grade~B, and the lower third form Grade~C. With $R_c = \overline{CF}^{\max}_c - \overline{CF}^{\min}_c$,
\begin{align}
\text{Grade~A (high)}   &: \quad \overline{CF}_i \in \big[\,\overline{CF}^{\max}_c - R_c/3,\ \ \overline{CF}^{\max}_c\,\big], \\
\text{Grade~B (medium)} &: \quad \overline{CF}_i \in \big[\,\overline{CF}^{\min}_c + R_c/3,\ \ \overline{CF}^{\max}_c - R_c/3\,\big), \\
\text{Grade~C (low)}    &: \quad \overline{CF}_i \in \big[\,\overline{CF}^{\min}_c,\ \ \overline{CF}^{\min}_c + R_c/3\,\big),
\end{align}
for every cell $i$ inside or intersecting country $c$, where $\overline{CF}_i$ is the cell's mean capacity factor over 1941--2025. The split is by resource quality, not by area or capacity, so a country with a few exceptionally good sites has a small Grade~A. Grade geometries are unions of the corresponding grid cells, and membership follows from the full 85-year mean to keep grades consistent across years (Fig.~\ref{fig:resource_grade_classification}). In the system model, the three grades of a country are modelled as separate generators, each with its own profile and installable potential.

\subsection{Technology parameters and wake losses}
\label{sec:supp_tech_wake}

\subsubsection{Solar tracking}
 The HSAT configuration uses the same silicon module as the baseline, but the panels rotate around a horizontal north--south axis to follow the sun from east to west, so the surface orientation and tilt are determined at each time step rather than fixed. Because tracking requires wider row spacing to limit self-shading, the maximum installation density is reduced from \num{5.1} to \qty{4.43}{\mega\watt\per\square\kilo\meter}.

 \subsubsection{Wind turbines}
\begin{figure}[htbp]
  \centering
  \includegraphics[width=0.89\linewidth]{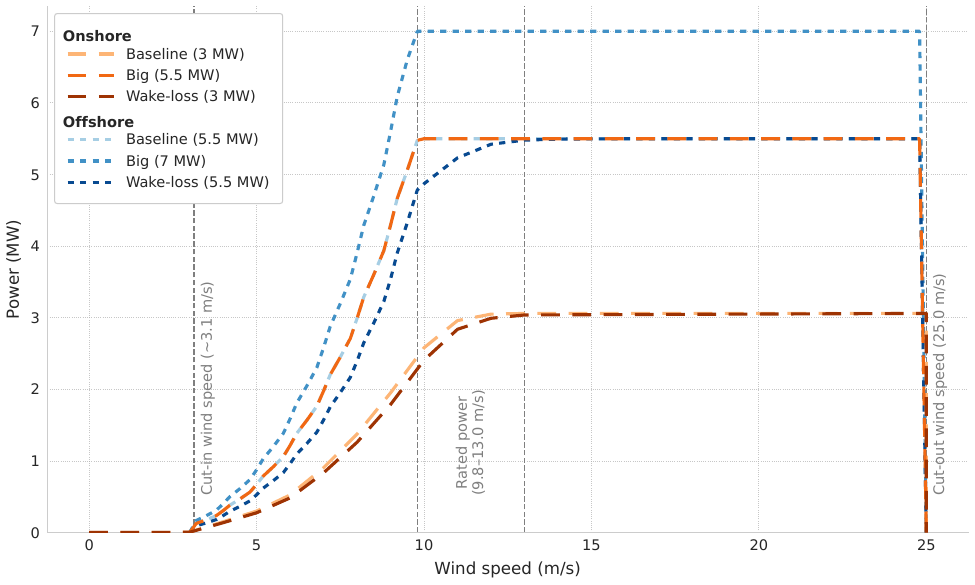}
  \caption{Power curves of the wind turbine models for onshore (orange) and offshore (blue) wind. Onshore, the baseline \qty{3}{\mega\watt} Vestas V112 (\qty{305}{\watt\per\square\meter}) is replaced in the larger-turbine configuration by the NREL 2020 ATB Reference \qty{5.5}{\mega\watt} turbine (\qty{229}{\watt\per\square\meter}), which also serves as the offshore baseline; offshore, it is replaced by the NREL 2020 ATB Reference \qty{7}{\mega\watt} turbine (\qty{223}{\watt\per\square\meter}). The darkest curves show the wake-modified power curves of the baseline turbines (Sec.~\ref{sec:wake-losses}), which reduce output below rated wind speed and shift rated output to higher wind speeds. Vertical dashed lines mark the cut-in (${\sim}$\qty{3.1}{\meter\per\second}) and cut-out (\qty{25}{\meter\per\second}) wind speeds and the range of rated wind speeds (\qtyrange{9.8}{13.0}{\meter\per\second}).}
  \label{fig:wind_turbine_models}
\end{figure}

 The baseline uses the Vestas V112 (\qty{3}{\mega\watt}, \qty{112}{m} rotor) onshore and the NREL 2020 ATB Reference \qty{5.5}{\mega\watt} turbine (\qty{175}{m} rotor) offshore. In the larger-turbine configuration, the latter replaces the V112 onshore, and the NREL 2020 ATB Reference \qty{7}{\mega\watt} turbine (\qty{200}{m} rotor) replaces the \qty{5.5}{\mega\watt} turbine offshore. Figure~\ref{fig:wind_turbine_models} shows the power curves of all turbine models together with the wake-modified curves of the two baseline turbines.

\subsubsection{Wake losses} 
Since exact plant layouts are unknown, we assume a circular layout with a representative capacity density of \qty{4}{\mega\watt\per\square\kilo\meter} and \num{48} turbines per plant onshore, and \qty{8}{\mega\watt\per\square\kilo\meter} and \num{64} turbines per plant offshore, based on empirical estimates \cite{covey_empirical_2026}. These are local wind-farm densities and therefore exceed the country-average installation limits (Sec.~\ref{sec:baseline_config}). The wake-loss curve accounts for both near-farm and mesoscale losses using the downwind propagation technique, in which every turbine casts a cone-shaped shadow with a Gaussian velocity deficit within it. Wake losses are averaged over all wind directions and computed with PyWake v2.6.2 \cite{simutis_enhancing_2024}. Losses between neighbouring wind farms are not represented.

\subsection{Winter capacity factors}
\label{sec:supp_winter_cf}

\begin{figure}[htbp]
  \centering
  \includegraphics[width=0.89\linewidth]{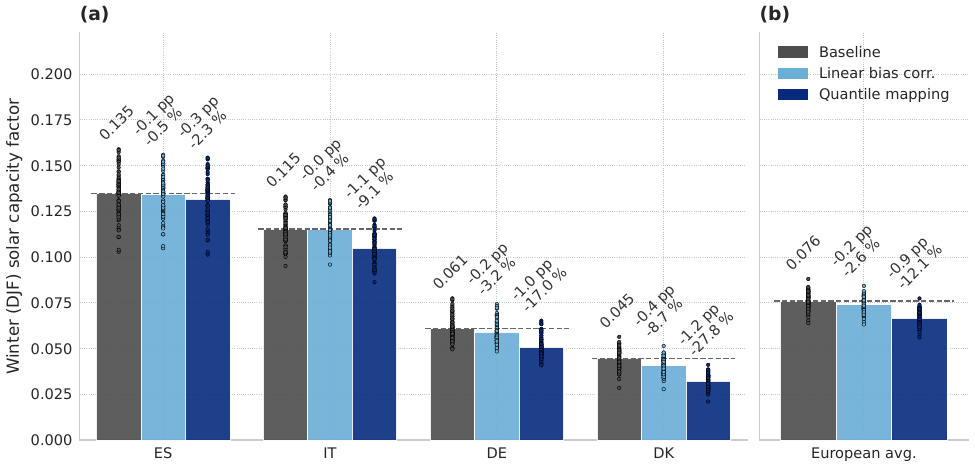}
  \caption{Comparison of average winter (December to February) solar capacity factors across the baseline, linear correction, and quantile mapping in (a) Spain (ES), Italy (IT), Germany (DE), and Denmark (DK); (b) load-weighted European average based on electricity demand in 2024. Bars indicate the mean winter capacity factor over the 84 winters, while dots represent individual winter values. The examined assumptions are compared against the baseline configuration.}
  \label{fig:solar-cf-winter}
\end{figure}

\begin{figure}[htbp]
  \centering
  \includegraphics[width=0.89\linewidth]{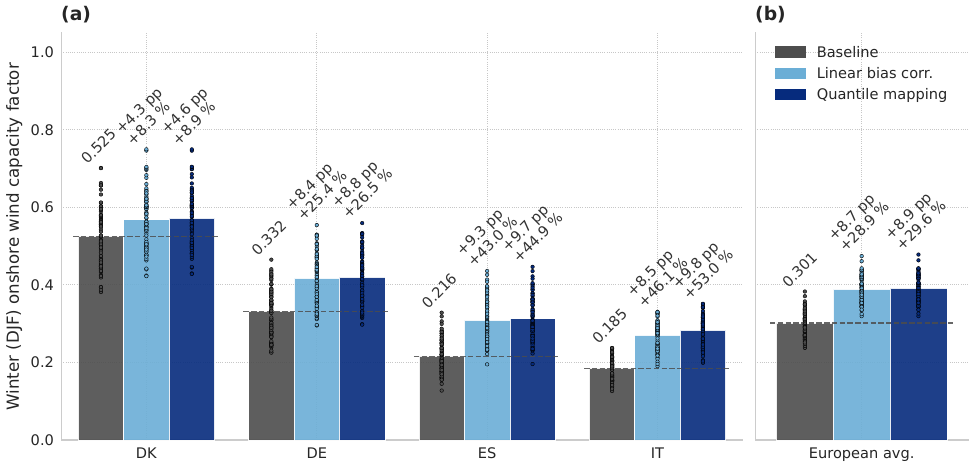}
  \caption{Comparison of average winter (December to February) onshore wind capacity factors across the baseline, linear correction, and quantile mapping in (a) Denmark (DK), Germany (DE), Spain (ES), and Italy (IT); (b) load-weighted European average based on electricity demand in 2024. Bars indicate the mean winter capacity factor over the 84 winters, while dots represent individual winter values. The examined assumptions are compared against the baseline configuration.}
  \label{fig:onshore-cf-winter}
\end{figure}

During winter, bias correction has a stronger effect on solar PV than over the full year, and quantile mapping reduces the CF more than linear correction (Fig.~\ref{fig:solar-cf-winter}). Although the absolute reduction is modest (up to \qty{1.2}{\pp} in Denmark), it amounts to a relative reduction of up to \qty{28}{\percent} owing to the low winter generation. For onshore wind, the uplift from bias correction is slightly larger in winter in absolute terms (up to \qty{+9.8}{\pp} in Italy) but smaller in relative terms, as winter CFs are higher (Fig.~\ref{fig:onshore-cf-winter}). The inter-annual variability of the solar baseline, measured as the standard deviation, is low at \qtyrange{3}{5}{\percent} of the mean annual CF (\qtyrange{0.5}{0.6}{\pp}), whereas individual years deviate considerably more (Fig.~\ref{fig:solar-cf-yearly}).

\subsection{Offshore wind annual and winter capacity factors}
\label{sec:supp_offshore_cf}

\begin{figure}[htbp]
  \centering
  \includegraphics[width=0.89\linewidth]{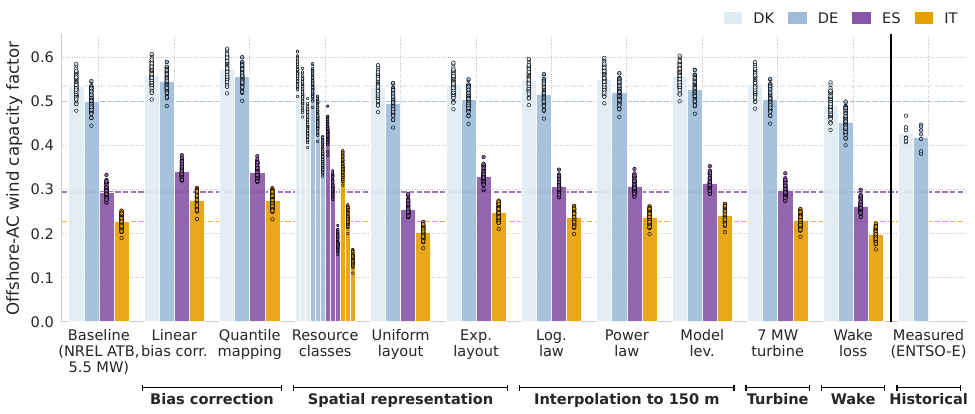}
  \caption{Mean annual offshore wind capacity factors across modelling assumptions in Denmark (DK), Germany (DE), Spain (ES), and Italy (IT). Bars show the mean over the 85 weather years, dots the individual annual values, and dashed lines the baseline mean of each country. For resource grades, the three bars per country show Grades~A to C from left to right. Configurations are grouped as in Fig.~\ref{fig:onshore-cf-yearly}. Historical values for 2015--2022 from Nayak et al.~\cite{nayak_validation_2025} (available for DK and DE only) are shown for context only, as the modelled capacity layout and turbines are not matched to the historical fleet.}
  \label{fig:offshore-cf-yearly}
\end{figure}

\begin{figure}[htbp]
  \centering
  \includegraphics[width=0.89\linewidth]{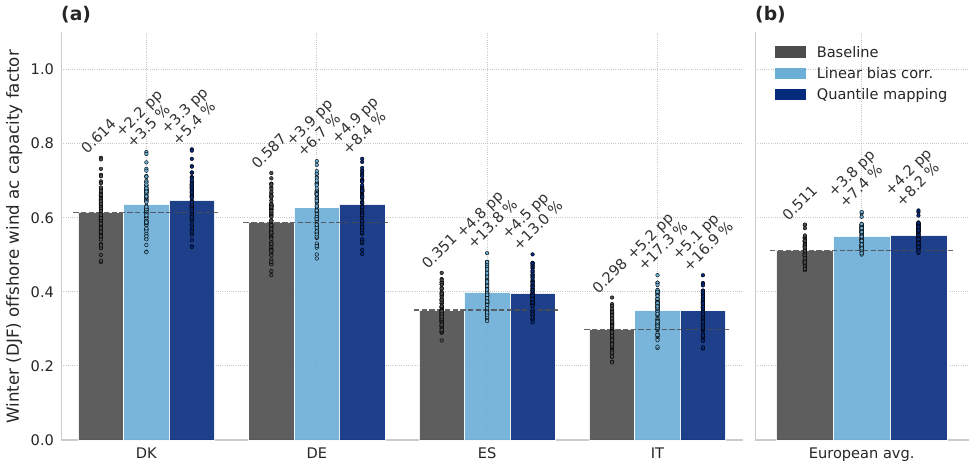}
  \caption{Comparison of average winter (December to February) offshore wind capacity factors across the baseline, linear correction, and quantile mapping in (a) Denmark (DK), Germany (DE), Spain (ES), and Italy (IT); (b) load-weighted European average based on electricity demand in 2024. Bars indicate the mean winter capacity factor over the 84 winters, while dots represent individual winter values. The examined assumptions are compared against the baseline configuration.}
  \label{fig:offshore-cf-winter}
\end{figure}

Offshore wind responds to the modelling assumptions similarly to onshore wind, but with a smaller spread, placing its overall sensitivity between that of solar PV and onshore wind (Fig.~\ref{fig:offshore-cf-yearly}). Bias correction raises mean CF by \qtyrange{2.5}{5.7}{\pp} (\qtyrange{5}{21}{\percent}), which in Spain and Italy is roughly half the onshore uplift, with quantile mapping exceeding linear correction by \qtyrange{1.2}{1.3}{\pp} in Denmark and Germany. In winter, the uplift is of similar absolute size (\qtyrange{2.2}{5.2}{\pp}) but smaller in relative terms (\qtyrange{3.5}{17}{\percent}), as winter CFs are higher (Fig.~\ref{fig:offshore-cf-winter}). As for onshore wind, the spatial layout mainly matters in countries with heterogeneous resources, shifting mean CF by up to \qty{3.9}{\pp} in Spain and Italy, but by less than \qty{0.5}{\pp} in Denmark and Germany. Raising the hub height from \num{120} to \qty{150}{\meter} increases CF by only \qtyrange{1.0}{2.8}{\pp}, since the change in height is smaller than onshore and the wind profile over the smooth sea surface is flatter, resulting in higher wind speeds already at the baseline height. The larger \qty{7}{\mega\watt} turbine has a negligible effect (\qtyrange{0.3}{0.5}{\pp}). In contrast, wake losses matter more for offshore than for onshore wind, reducing mean CF by \qtyrange{2.9}{4.7}{\pp} (\qtyrange{8}{13}{\percent}), which is twice the onshore reduction (\qtyrange{0.8}{2.2}{\pp}), reflecting the denser plant layout assumed offshore.

\subsection{Wind speed bias correction factors and capacity factor distributions}
\label{sec:supp_bias_cf_dist}

The quantile-specific wind speed correction factors (Fig.~\ref{fig:bias_correction_era5_gwa4_vertical}b--d) differ clearly from each other. The high wind speeds (95th percentile; Fig.~\ref{fig:bias_correction_era5_gwa4_vertical}d) generally receive a strong upward correction, stronger than that for median wind speeds (Fig.~\ref{fig:bias_correction_era5_gwa4_vertical}b). High correction factors exceeding 2.5 can be observed in complex terrain such as the Norwegian coast and the Balkans. The filtering criteria (Sec.~\ref{sec:qm_correction}) exclude large parts of the Alps and Pyrenees from quantile mapping.

Despite these differences, both bias correction techniques result in almost the same distribution of CF per country, with only slight variations at the tails (Fig.~\ref{fig:cf_distribution_with_reference}). Much of the quantile-specific behaviour is thus dampened when wind speeds are aggregated spatially and converted to power output. In Spain and Italy (Fig.~\ref{fig:cf_distribution_with_reference}c,d), the bias-corrected distribution is shifted much closer to the historical one \cite{nayak_validation_2025}, since the baseline strongly overestimates the occurrence of low-generation hours (CF\,$<$\,0.1). The opposite is seen for Germany and Denmark (Fig.~\ref{fig:cf_distribution_with_reference}a,b). Consistent with the mean CF comparison, the German baseline reproduces the historical distribution reasonably well, while bias correction adds some deviations. Denmark shows the largest deviation: the modelled distribution has a pronounced peak of very high capacity factors (CF\,$\approx$\,1), which could be explained by the relatively small country area, and underestimates the hours of low generation relative to the historical distribution. This effect becomes larger after bias correction with GWA v4 (Fig.~\ref{fig:cf_distribution_with_reference}a), which reports higher wind speeds over Denmark than GWA v2 (Figs.~\ref{fig:bias_factor_gwa2_gwa4_era5} and~\ref{fig:bias_factor_gwa2_gwa4}).

\subsection{Bias correction versus calibrated power-curve smoothing}
\label{sec:supp_gaussian_smoothing}

Gaussian smoothing of the turbine power curve has been applied to represent the behaviour of wind power fleets across large areas \cite{andresen_validation_2015,staffell_using_2016}. The parameters of the Gaussian function are determined by minimising the divergence between the modelled and historical duration curves of the country-wide capacity factor time series. In practice, this allows the smoothing to account for several effects simultaneously: a potential wind speed bias in the weather input data, the spatial layout of the capacity, the aggregated behaviour of the turbine fleet (including the unavailability of individual units and the heterogeneity of hub heights and power curves across turbines), and the calculation of hourly capacity factors from instantaneous wind speed data. However, this approach is highly sensitive to the quality of the historical data used for calibration, and it does not reveal which of these effects drives the improvement. By evaluating each modelling assumption individually, we show to what extent the agreement between modelled and historical duration curves can be improved by bias-correcting the wind speed alone (Fig.~\ref{fig:cf_distribution_with_reference}; Sec.~\ref{sec:supp_bias_cf_dist}).

\subsection{Onshore wind energy droughts}
\label{sec:supp_drought_onwind}

\begin{figure}[htbp]
    \centering
    \includegraphics[width=0.89\linewidth]{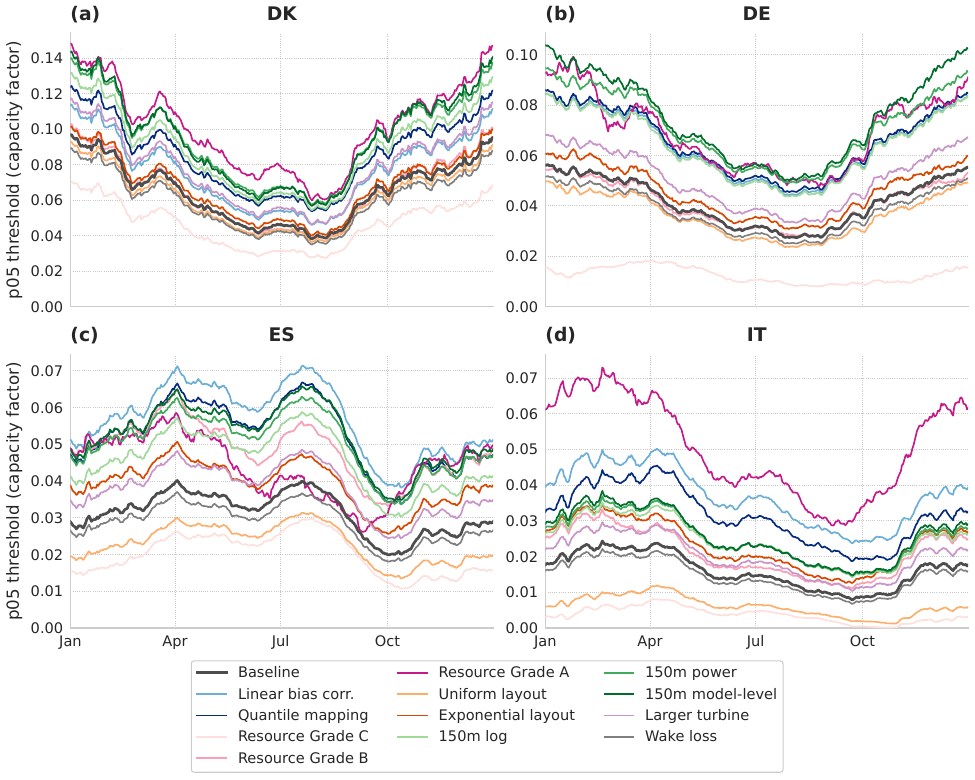}
    \caption{Smoothed day-of-year p5 capacity factor thresholds for all onshore-wind sensitivity configurations, per country. In DK and DE, the configurations largely shift the threshold vertically while preserving the common seasonal cycle (winter maximum, summer minimum). In ES, however, several configurations alter the seasonal shape itself: the highest resource grade peaks in early spring and falls below the bias-corrected series in summer, when a secondary maximum emerges. These configurations thus change the distribution of low-wind conditions over the year rather than merely rescaling it.}
    \label{fig:drought_threshold_onwind_doy_countries}
\end{figure}

Droughts are identified for each configuration from its individual day-of-year p5 threshold (Fig.~\ref{fig:drought_threshold_onwind_doy_countries}; Sec.~\ref{sec:droughts}). In the baseline, averaged across the four focus countries, onshore wind droughts occur 8.1 times per year with a mean duration of 1.87 days. Across all combinations of assumption, country, and season, frequency and duration deviate from the baseline by only about \qty{2.4}{\percent} on average (solar PV: about \qty{3}{\percent}, Sec.~\ref{sec:supp_drought_solar}; offshore wind: about \qty{2}{\percent}, Sec.~\ref{sec:supp_drought_offwind}).

\begin{figure}[htbp]
  \centering
  \includegraphics[width=0.89\linewidth]{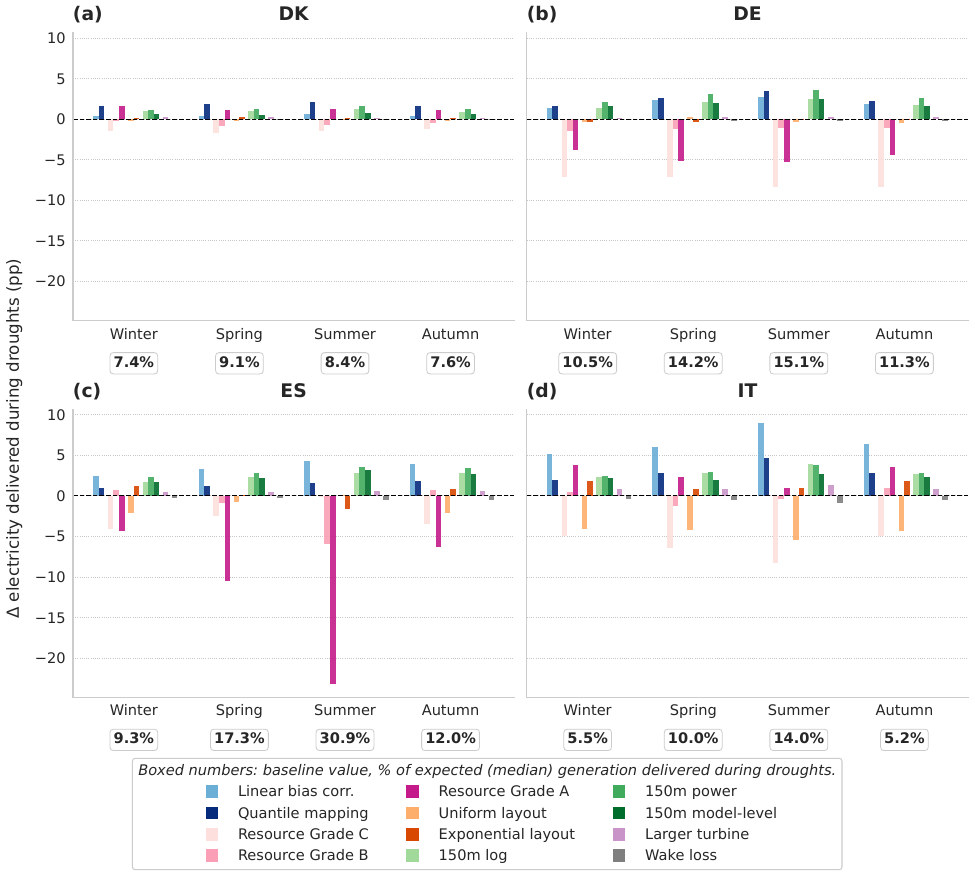}
  \caption{Change in the percentage of expected (own day-of-year median) electricity delivered during each configuration's own identified onshore-wind drought periods (individually calibrated p5 threshold per configuration; Eq.~\eqref{eq:delivered}), relative to the onshore-wind baseline, for (a) Denmark, (b) Germany, (c) Spain, and (d) Italy. Values are shown by season, in percentage points; positive values indicate more electricity delivered than in the baseline, i.e.\ a smaller drought-related shortfall. Numbers below the x-axis give the baseline's own value for each season: the percentage of its expected output delivered during its own droughts.}
  \label{fig:drought_delivered_onwind_seasons_countries}
\end{figure}

Figure~\ref{fig:drought_delivered_onwind_seasons_countries} extends the winter results of the main text (Fig.~\ref{fig:drought_delivered_onwind_winter_countries}) to all seasons. The direction of the effects is largely consistent across seasons; their magnitude is small in Denmark but substantially larger in Germany, Spain, and Italy. The choice of bias correction method matters primarily in complex terrain, where linear correction increases the delivered share more than quantile mapping does. Denmark is a clear exception, as the 5th percentile of the quantile-mapping correction factor exceeds the linear one there (Fig.~\ref{fig:bias_correction_era5_gwa4_vertical}a,c). A larger turbine slightly reduces shortfalls, and wake losses reduce the delivered share by less than \qty{1}{\pp} in all countries and seasons, in contrast to offshore wind (Fig.~\ref{fig:drought_delivered_offwind-ac_seasons_countries}). In the load-weighted European average (Fig.~\ref{fig:drought_delivered_onwind_seasons_europe_avg}), Resource Grade~C has by far the largest single effect, and the uniform layout lowers the delivered share slightly.

\subsection{Solar energy droughts}
\label{sec:supp_drought_solar}

\begin{figure}[htbp]
  \centering
  \includegraphics[width=0.89\linewidth]{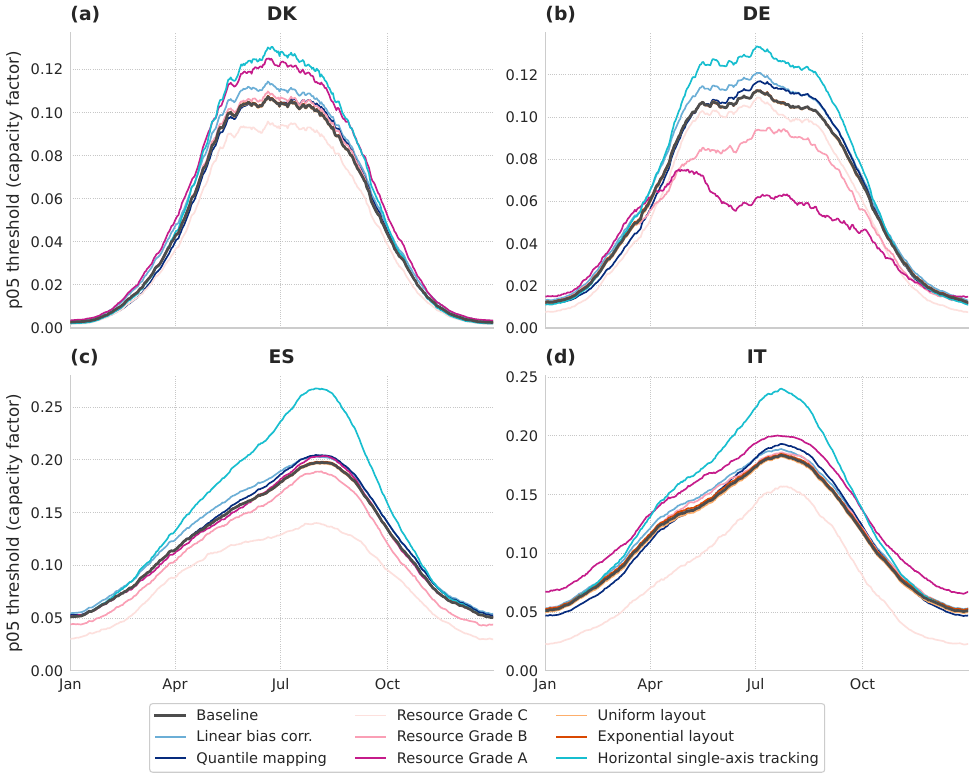}
  \caption{Smoothed day-of-year p5 capacity factor thresholds for all solar PV configurations, for (a)~Denmark, (b)~Germany, (c)~Spain, and (d)~Italy. The thresholds follow the strong seasonal cycle of solar generation and approach zero in winter in Denmark and Germany. HSAT raises the summer thresholds most, while in Germany Resource Grade~A lies well below the other configurations in summer.}
  \label{fig:drought_threshold_solar_doy_countries}
\end{figure}

\begin{figure}[htbp]
  \centering
  \includegraphics[width=0.89\linewidth]{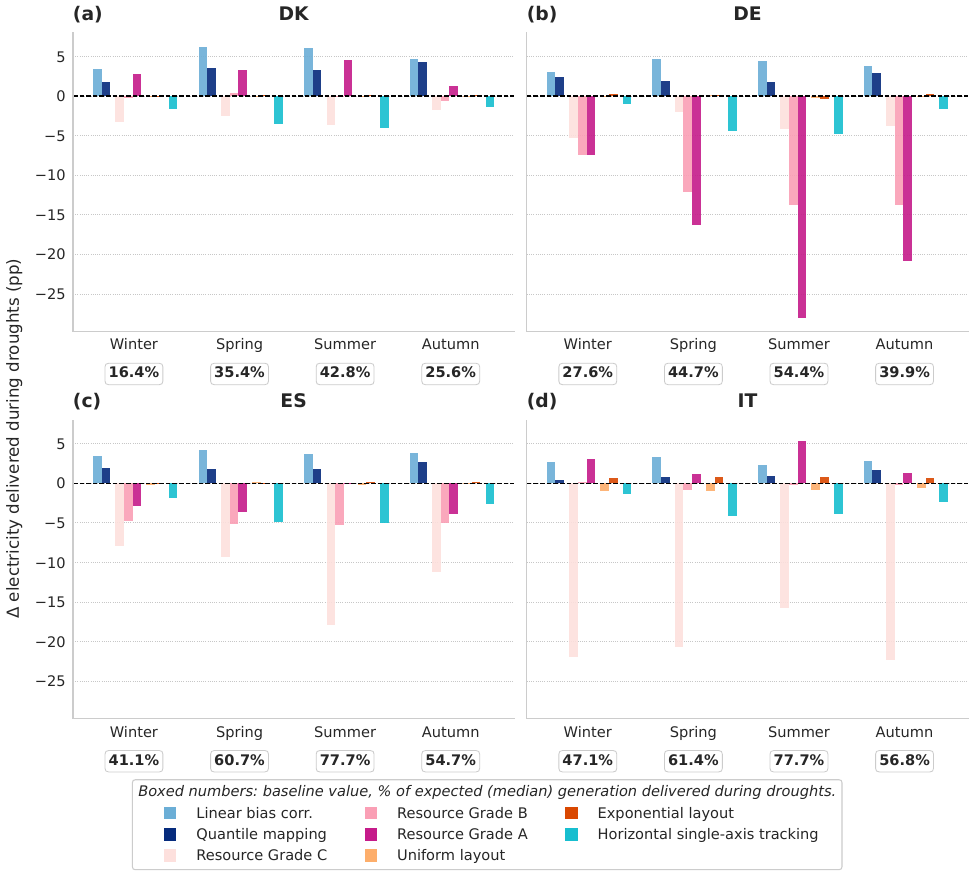}
  \caption{Change in the percentage of expected (own day-of-year median) electricity delivered during each configuration's own identified solar drought periods (individually calibrated p5 threshold per configuration; Eq.~\eqref{eq:delivered}), relative to the solar baseline, for (a) Denmark, (b) Germany, (c) Spain, and (d) Italy. Values are shown by season, in percentage points; positive values indicate more electricity delivered than in the baseline, i.e.\ a smaller drought-related shortfall. Numbers below the x-axis give the baseline's own value for each season: the percentage of its expected output delivered during its own droughts.}
  \label{fig:drought_delivered_solar_seasons_countries}
\end{figure}

Solar droughts are identified analogously (Fig.~\ref{fig:drought_threshold_solar_doy_countries}), but from daily CFs to avoid night-time hours (Sec.~\ref{sec:droughts}). Baseline solar droughts are nearly twice as frequent as droughts of either wind technology (15.4 events per year vs.\ 8.1 for both onshore and offshore wind) but shorter (mean length 1.46 days vs.\ 1.87 and 1.82 days). Compared with wind, solar shows a higher sensitivity to modelling assumptions. On average, frequency and length deviate only slightly from the baseline (by about \qty{3}{\percent}), but by up to \qty{+30}{\percent}/\qty{-19}{\percent} for Resource Grade~C in individual country-seasons, which is larger than the maximum deviations of about \qty{23}{\percent} and \qty{15}{\percent} for onshore and offshore wind, respectively.
Figure~\ref{fig:drought_delivered_solar_seasons_countries} shows that solar droughts are also shallower in relative terms, thus delivering a higher share of electricity even in critical periods. The baseline delivers \qtyrange{16}{78}{\percent} of expected output during its own droughts, versus \qtyrange{5}{31}{\percent} for onshore wind and \qtyrange{8}{31}{\percent} for offshore wind. Bias-correcting solar irradiance consistently increases the delivered share during droughts, while Resource Grade~C reduces it, with the effect increasing sharply from north to south. It is small in Denmark, modest in Germany, and much larger in Spain and Italy, reaching up to about \num{18} and \qty{22}{\pp}, respectively. On average across the four countries, Resource Grade~C's effect is larger for solar than for wind (\qty{9.6}{\pp} vs.\ \num{4.5} for onshore wind and \num{3.9} for offshore wind). HSAT, which on average increases the CF, consistently reduces the delivered share during droughts (by up to about \qty{5}{\pp}). Notably, in Germany all individual resource grades reduce the delivered share during droughts.

\subsection{Offshore wind energy droughts}
\label{sec:supp_drought_offwind}

\begin{figure}[htbp]
  \centering
  \includegraphics[width=0.89\linewidth]{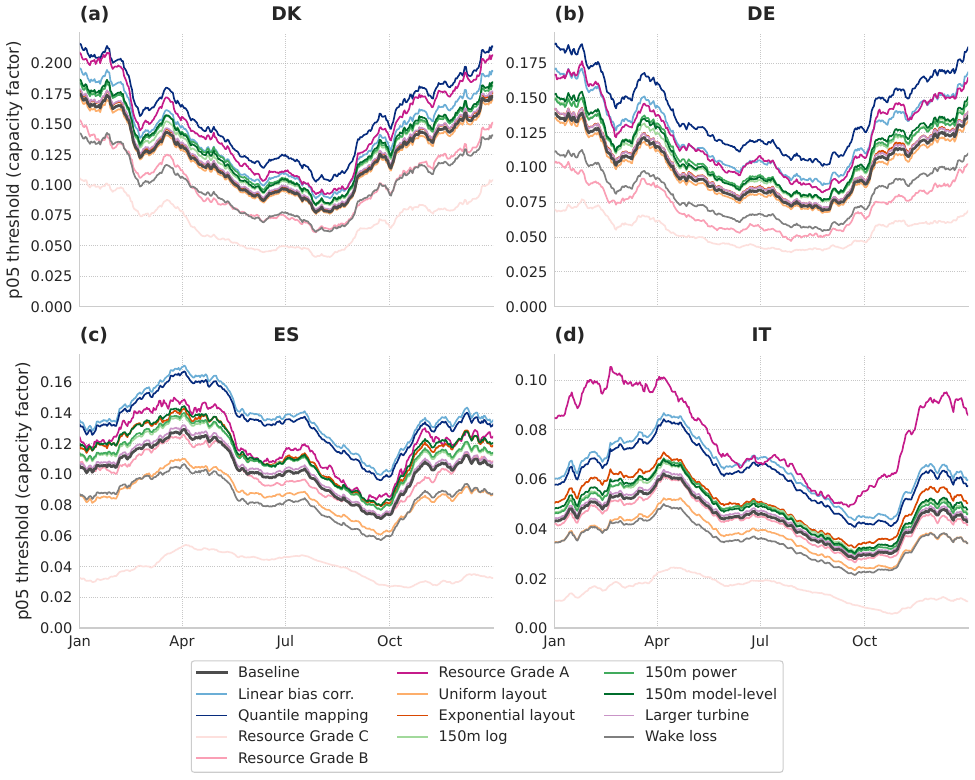}
  \caption{Smoothed day-of-year p5 capacity factor thresholds for all offshore wind (AC) configurations, for (a)~Denmark, (b)~Germany, (c)~Spain, and (d)~Italy. All configurations show a winter maximum and a summer minimum. Bias correction raises the thresholds in all countries and Resource Grade~A strongly in Italy, while Resource Grade~C lowers them most.}
  \label{fig:drought_threshold_offwind-ac_doy_countries}
\end{figure}

\begin{figure}[htbp]
  \centering
  \includegraphics[width=0.89\linewidth]{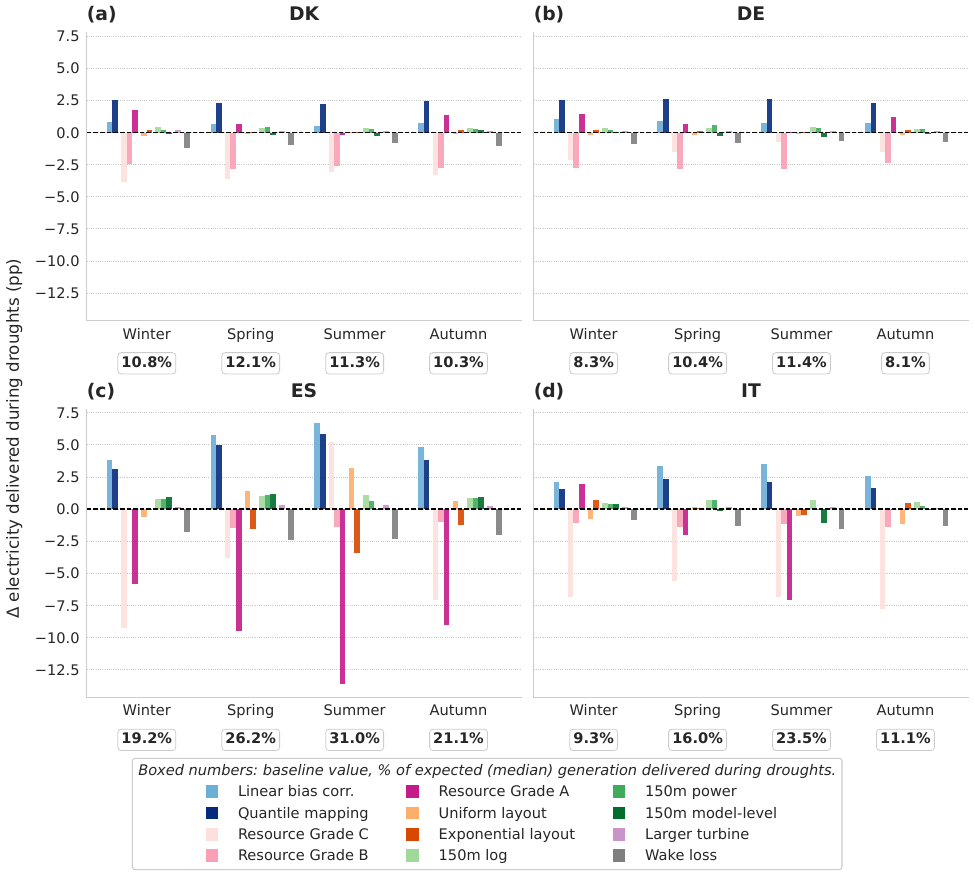}
  \caption{Change in the percentage of expected (own day-of-year median) electricity delivered during each configuration's own identified offshore-wind drought periods (individually calibrated p5 threshold per configuration; Eq.~\eqref{eq:delivered}), relative to the offshore-wind baseline, for (a) Denmark, (b) Germany, (c) Spain, and (d) Italy. Values are shown by season, in percentage points; positive values indicate more electricity delivered than in the baseline, i.e.\ a smaller drought-related shortfall. Numbers below the x-axis give the baseline's own value for each season: the percentage of its expected output delivered during its own droughts.}
  \label{fig:drought_delivered_offwind-ac_seasons_countries}
\end{figure}

Offshore wind droughts are identified in the same way as onshore wind droughts (Fig.~\ref{fig:drought_threshold_offwind-ac_doy_countries}). The frequency and duration of droughts do not change much across different configurations and are close to those of onshore wind. Relative to the baseline (8.1 events per year, mean length 1.82 days, averaged across the four focus countries), frequency and length deviate by only about \qty{2}{\percent} on average across all assumption/country/season combinations, and by no more than \qty{15}{\percent} even in the most extreme individual country-season case (Resource Grade~C, Spain, autumn). Figure~\ref{fig:drought_delivered_offwind-ac_seasons_countries} shows the effect of the modelling assumptions on electricity delivered during identified drought periods. Bias correction (linear and quantile mapping) consistently increases the delivered share relative to the baseline. By contrast, the resource grade configurations mostly decrease it, most strongly for Resource Grade~C, which reduces the delivered share by \qty{3.9}{\pp} on average across the four countries. The baseline itself delivers \qtyrange{8}{31}{\percent} of its expected output during its own droughts, depending on country and season, reaching its highest value (\qty{31}{\percent}) in Spain during summer, which is virtually identical to onshore wind's own maximum there. The comparison with onshore wind, however, is not uniform. During droughts, raising the hub height increases generation more for onshore than for offshore wind. In contrast, wake losses reduce the delivered share considerably more offshore, by up to about \qty{2.4}{\pp} in Spain compared with less than \qty{1}{\pp} onshore. For bias correction, the changes are more heterogeneous and country-dependent. 
In the load-weighted European average, linear correction and quantile mapping change positions relative to onshore wind (Fig.~\ref{fig:drought_delivered_offwind-ac_seasons_europe_avg}).

\subsection{European electricity shortfall during energy droughts and sensitivity to the drought threshold}
\label{sec:supp_threshold}

\begin{figure}[htbp]
  \centering
  \includegraphics[width=0.89\linewidth]{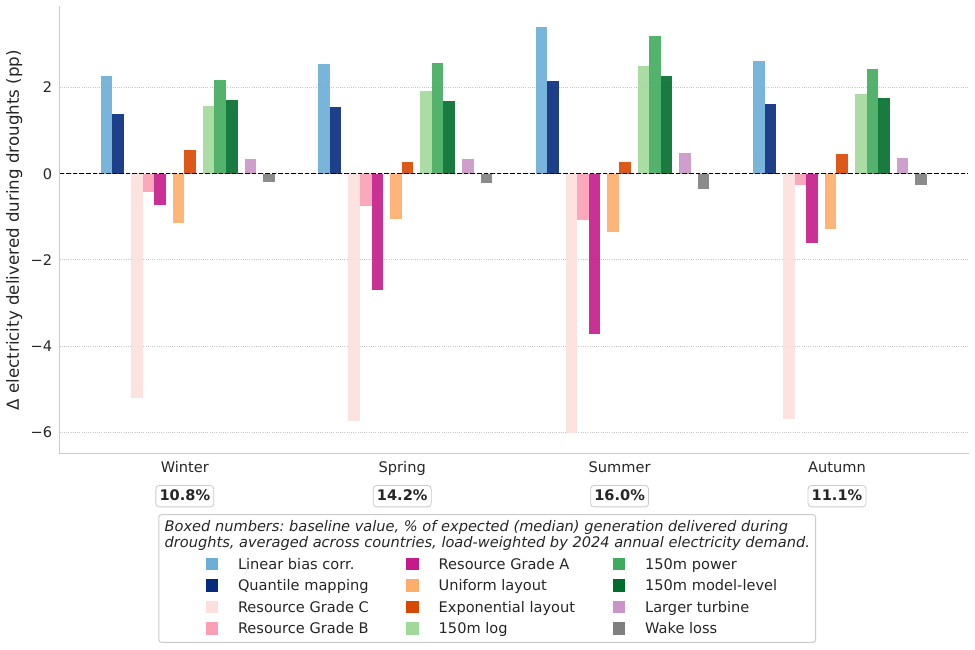}
  \caption{\textbf{Onshore wind, load-weighted European average, p5 threshold.} Change in the delivered share during droughts (load-weighted average of $\bar{\delta}_c$ across countries; Eq.~\eqref{eq:delivered}), i.e.\ the percentage of expected electricity (own day-of-year median) delivered during each configuration's own droughts, relative to the onshore-wind baseline, by season and in percentage points. Droughts are identified separately for each configuration from its own p5 threshold. Countries are weighted by their 2024 annual electricity demand; Albania, Bosnia and Herzegovina, Montenegro, North Macedonia, Slovenia, and Switzerland are excluded owing to their very low onshore wind generation. Positive values indicate more electricity delivered than in the baseline, i.e.\ a smaller drought-related shortfall. Numbers below the x-axis give the baseline's load-weighted delivered share for each season.}
  \label{fig:drought_delivered_onwind_seasons_europe_avg}
\end{figure}

\begin{figure}[htbp]
  \centering
  \includegraphics[width=0.7\linewidth]{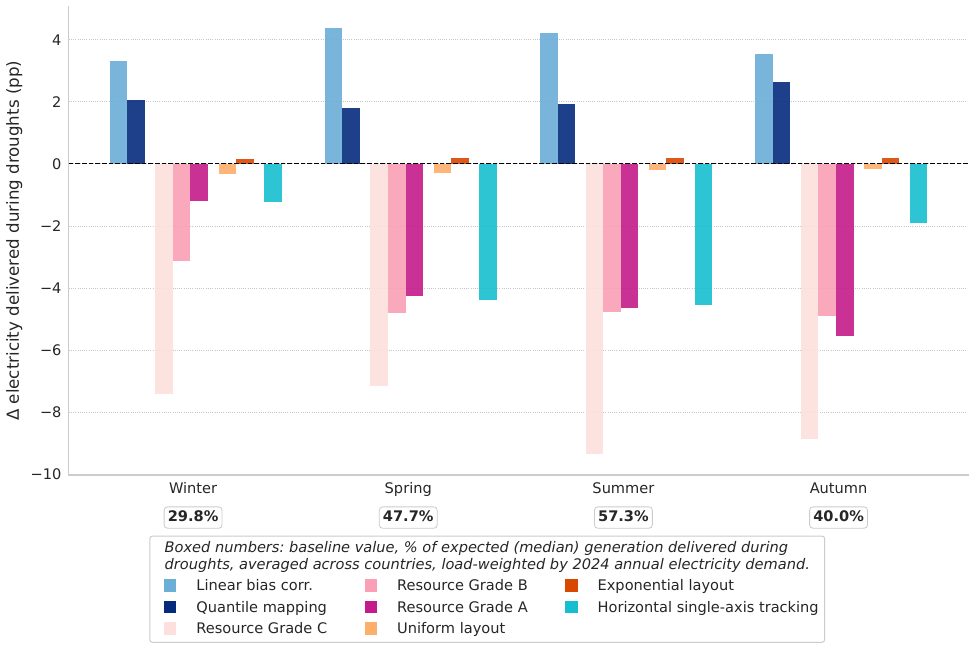}
  \caption{\textbf{Solar PV, load-weighted European average, p5 threshold.} As Fig.~\ref{fig:drought_delivered_onwind_seasons_europe_avg}, but for solar PV droughts relative to the solar baseline. All countries are included.}
  \label{fig:drought_delivered_solar_seasons_europe_avg}
\end{figure}

\begin{figure}[htbp]
  \centering
  \includegraphics[width=0.7\linewidth]{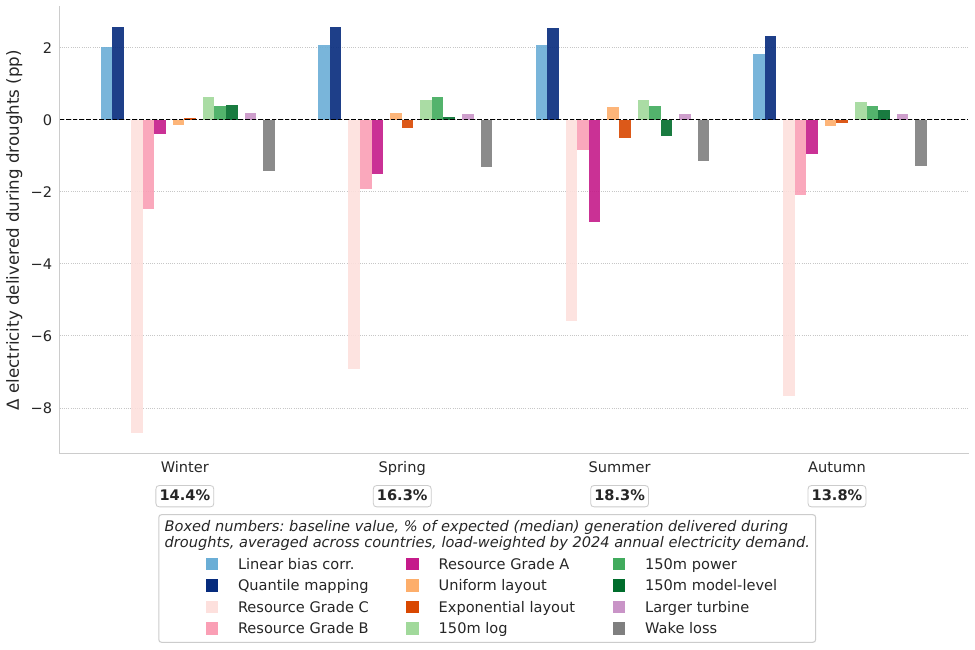}
  \caption{\textbf{Offshore wind, load-weighted European average, p5 threshold.} As Fig.~\ref{fig:drought_delivered_onwind_seasons_europe_avg}, but for offshore wind droughts relative to the offshore-wind baseline. All countries with offshore wind are included.}
  \label{fig:drought_delivered_offwind-ac_seasons_europe_avg}
\end{figure}

\begin{figure}[htbp]
  \centering
  \includegraphics[width=0.89\linewidth]{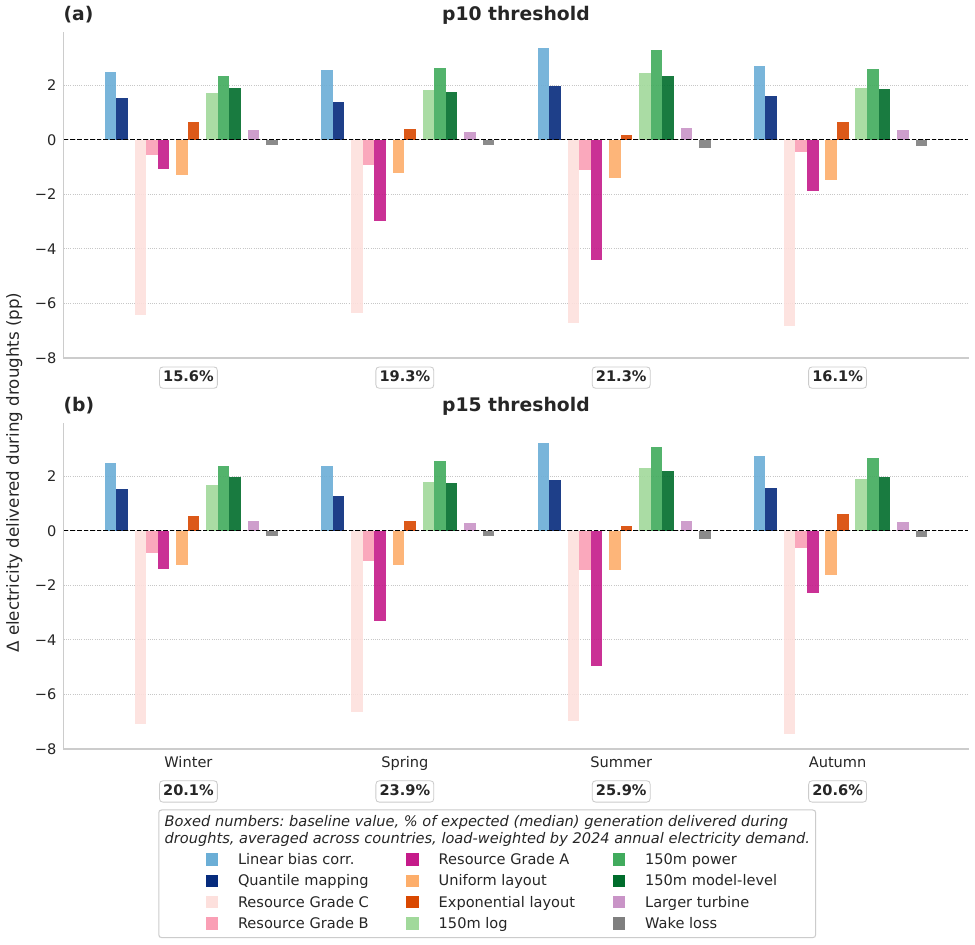}
  \caption{\textbf{Onshore wind, load-weighted European average, threshold sensitivity.} As Fig.~\ref{fig:drought_delivered_onwind_seasons_europe_avg}, but with droughts identified from each configuration's (a)~p10 and (b)~p15 threshold.}
  \label{fig:drought_delivered_onwind_europe_p10_p15}
\end{figure}

\begin{figure}[htbp]
  \centering
  \includegraphics[width=0.89\linewidth]{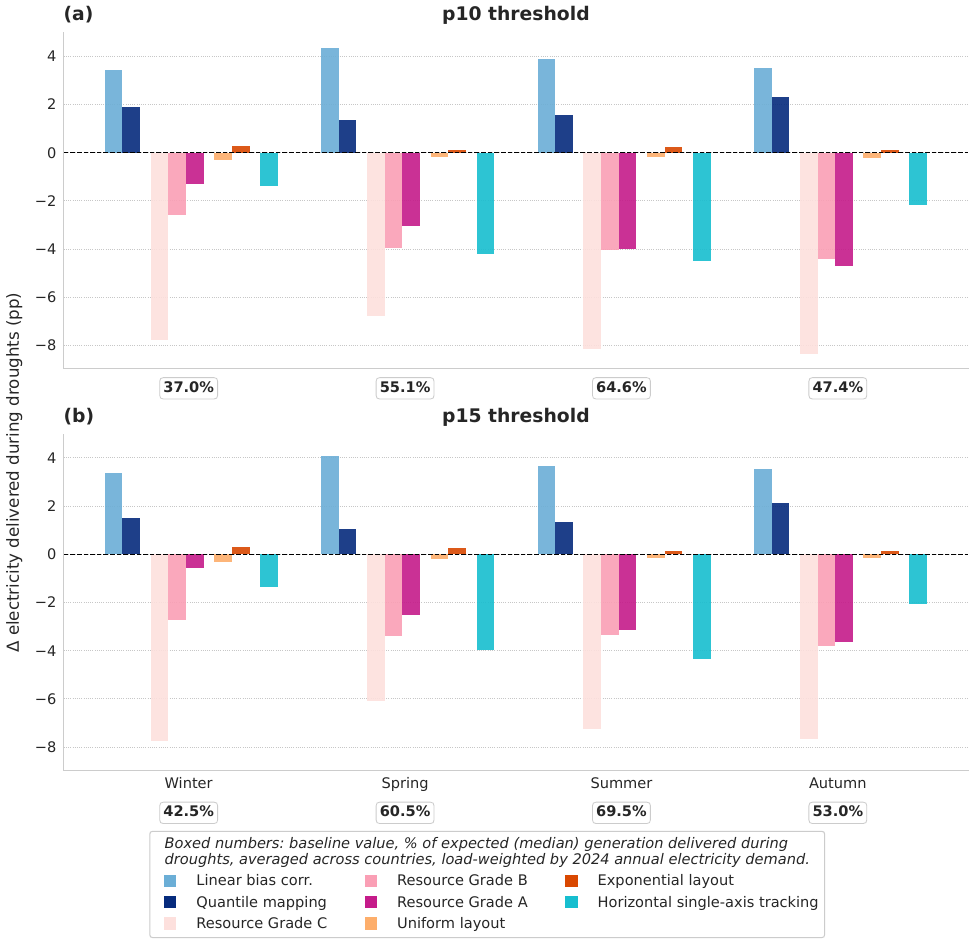}
  \caption{\textbf{Solar PV, load-weighted European average, threshold sensitivity.} As Fig.~\ref{fig:drought_delivered_solar_seasons_europe_avg}, but with droughts identified from each configuration's (a)~p10 and (b)~p15 threshold.}
  \label{fig:drought_delivered_solar_europe_p10_p15}
\end{figure}

\begin{figure}[htbp]
  \centering
  \includegraphics[width=0.89\linewidth]{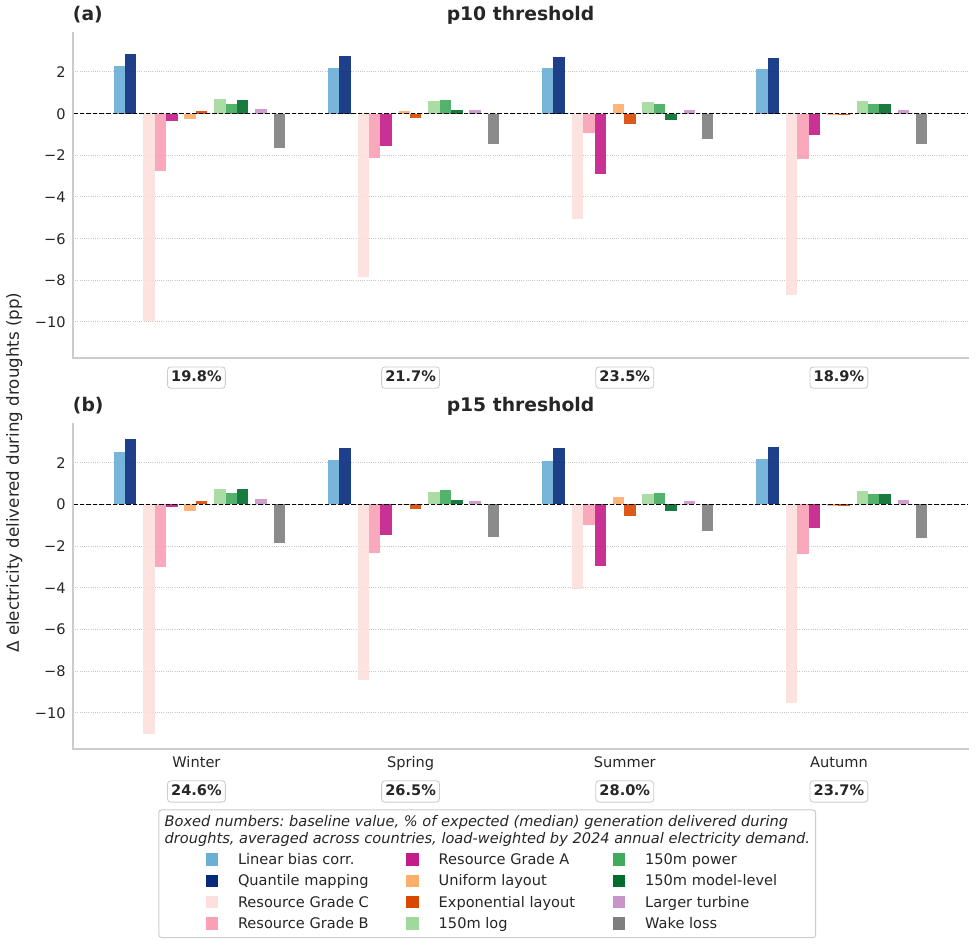}
  \caption{\textbf{Offshore wind, load-weighted European average, threshold sensitivity.} As Fig.~\ref{fig:drought_delivered_offwind-ac_seasons_europe_avg}, but with droughts identified from each configuration's (a)~p10 and (b)~p15 threshold.}
  \label{fig:drought_delivered_offwind-ac_europe_p10_p15}
\end{figure}

Figures~\ref{fig:drought_delivered_onwind_seasons_europe_avg}--\ref{fig:drought_delivered_offwind-ac_seasons_europe_avg} show the load-weighted European average of the delivered share during droughts for onshore wind, solar PV, and offshore wind, based on the p5 threshold used throughout the main text. Figures~\ref{fig:drought_delivered_onwind_europe_p10_p15}--\ref{fig:drought_delivered_offwind-ac_europe_p10_p15} repeat the analysis with droughts identified from the 10th (p10) and 15th (p15) percentile thresholds; all other identification rules are unchanged. The direction and ranking of the effects of the modelling choices are consistent across the three thresholds.

\subsubsection{Main threshold (p5)}

\subsubsection{Threshold sensitivity (p10 and p15)}

\subsection{Changes in the supply of the European energy system}

\begin{figure}[htbp]
  \centering
  \includegraphics[width=0.89\linewidth]{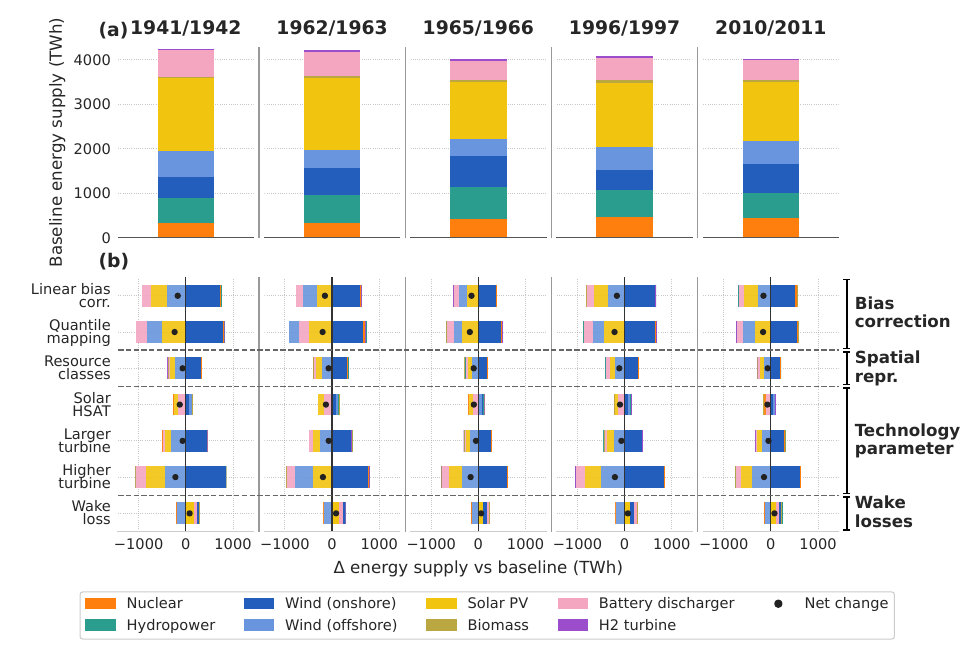}

  \caption{Total energy supply per weather year. (a) Baseline energy supply by technology for each of the five weather years, comprising all electricity fed into the nodes, including storage discharge. (b) Change in energy supply relative to the corresponding baseline for each modelling assumption; dots mark the net change.}

  \label{fig:energy_supply_system}
\end{figure}

Although the exogenous demand is identical across configurations (about \qty[group-minimum-digits=4]{3300}{TWh} per weather year), the total energy supply differs by up to \qty{5.6}{\percent} (\num{-238} to \qty{+85}{TWh}; Fig.~\ref{fig:energy_supply_system}). Supply here comprises all electricity fed into the nodes by generation and storage discharge and therefore exceeds demand. The differences stem mainly from battery storage. In the bias-corrected systems, onshore wind displaces solar PV, so less energy has to be shifted within the day, and battery discharge falls on average from \qty{507}{TWh} to \qty{358}{TWh} (linear) and \qty{312}{TWh} (quantile mapping). This is a capacity rather than a utilisation effect: under quantile mapping, battery power capacity falls by \qty{37}{\percent} and discharge by \qty{38}{\percent}, while the number of full cycles per year remains almost unchanged (301 versus 287). Wake losses act in the opposite direction, increasing solar PV and battery capacity and raising total supply by \qtyrange{60}{85}{TWh}. HSAT reduces the cycling of the remaining batteries the most (\qty{7}{\percent} fewer full cycles per year), as its broader daily generation profile reduces the midday surplus (Fig.~\ref{fig:diurnal_cycle_solar}). Transmission has little impact, since DC lines are modelled as lossless and AC line losses (about \qty{50}{TWh}) change by up to \qty{11.5}{TWh} between configurations.

\subsection{Identified system stress event overview}
\label{sec:supp_stress_events}

\begin{figure}[htbp]
  \centering
  \includegraphics[width=0.89\linewidth]{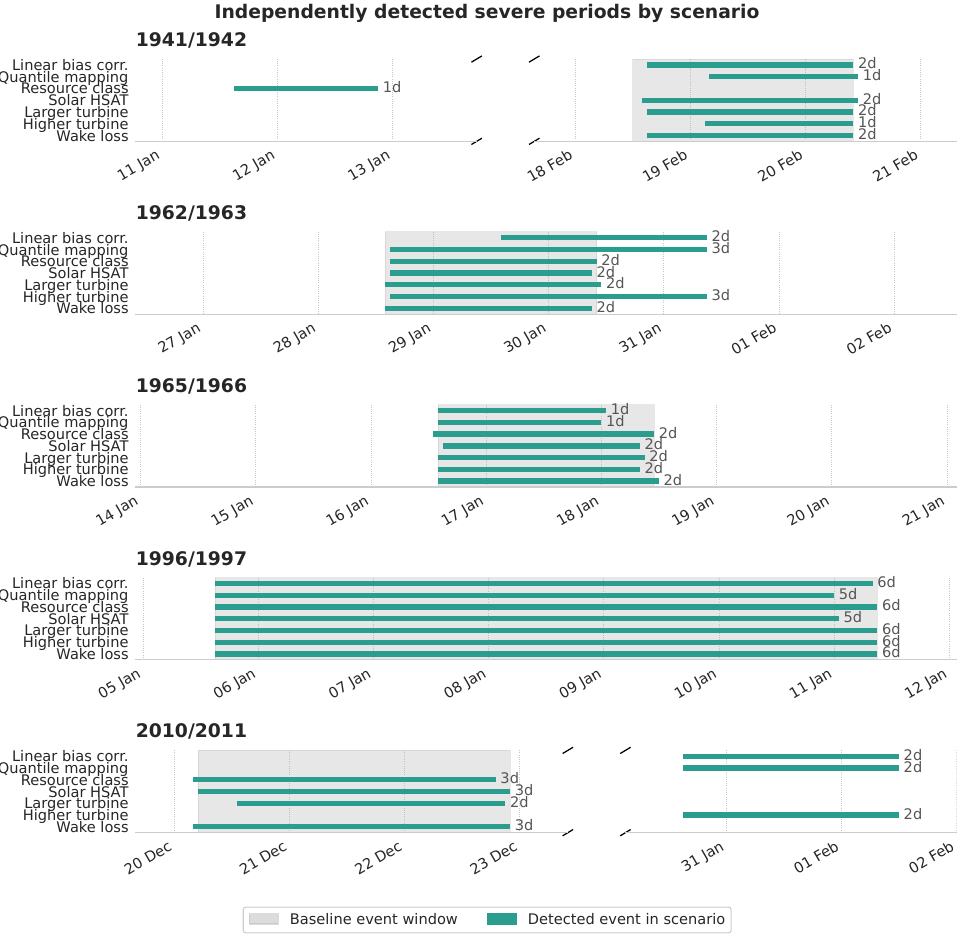}
  \caption{Independently detected severe periods by configuration for the five
    critical weather years. Grey shading marks the baseline event window; bars
    show the period detected in each configuration, annotated with its duration.
    Broken axes separate windows where configurations detect different periods.}
  \label{fig:stress_events_overview}
\end{figure}

Figure~\ref{fig:stress_events_overview} compares, for the five critical weather years, the most severe system stress period detected independently in each configuration against the baseline event window. The timing of stress events is largely robust to modelling choices (e.g.\ in 1962/63, 1965/66, and 1996/97). In these years, all configurations identify essentially the same period, with event durations differing by only a few days. The bias-corrected, \qty{150}{\meter} hub-height, and resource-grade configurations can reorder events. In 1941/42 and 2010/11, they shift the most severe period to a different window from the baseline's, although the original period still causes high stress.

\begin{figure}[htbp]
  \centering
  \includegraphics[width=0.89\linewidth]{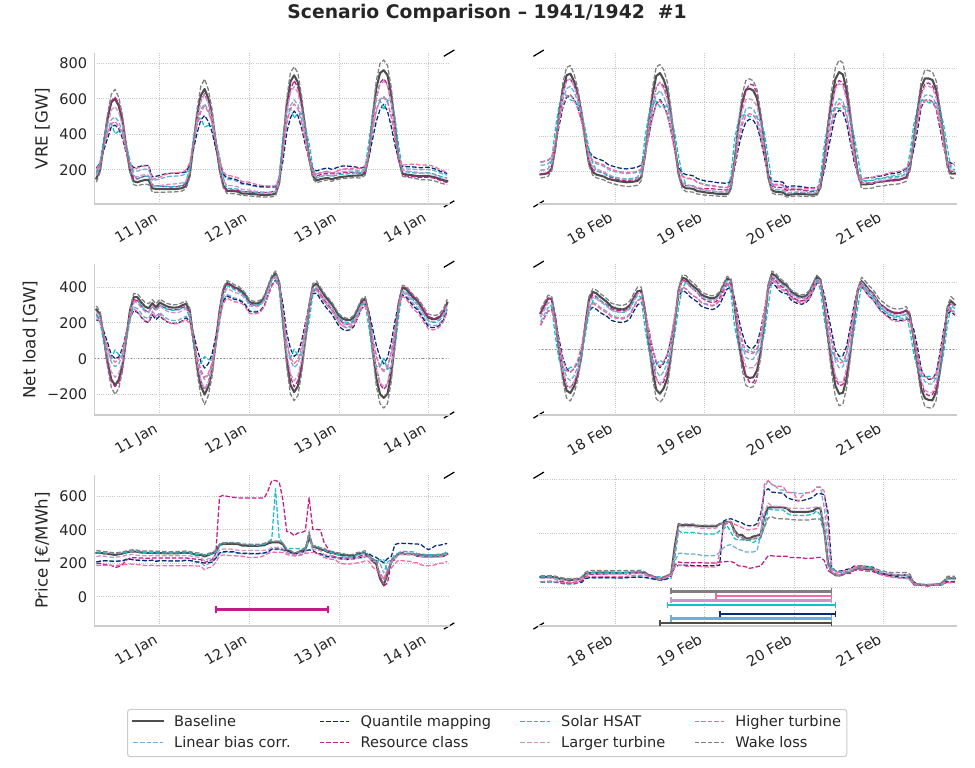}
  \caption{Stress event \#1 (weather year 1941/42): VRE generation (top), net load (middle), and electricity price (bottom) for the baseline and all modelling-choice configurations. Horizontal bars mark the severe period detected in each configuration; broken axes separate the two detected windows.}
  \label{fig:stress_event_1}
\end{figure}

\begin{figure}[htbp]
  \centering
  \includegraphics[width=0.89\linewidth]{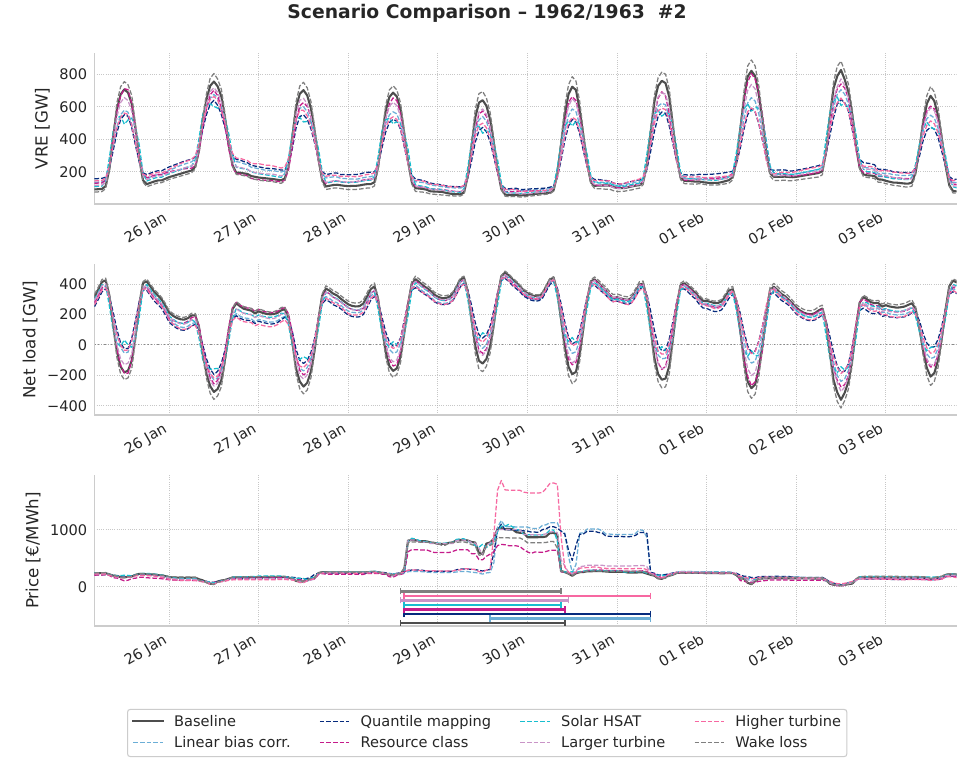}
  \caption{Stress event \#2 (weather year 1962/63): as
    Fig.~\ref{fig:stress_event_1}.}
  \label{fig:stress_event_2}
\end{figure}

\begin{figure}[htbp]
  \centering
  \includegraphics[width=0.89\linewidth]{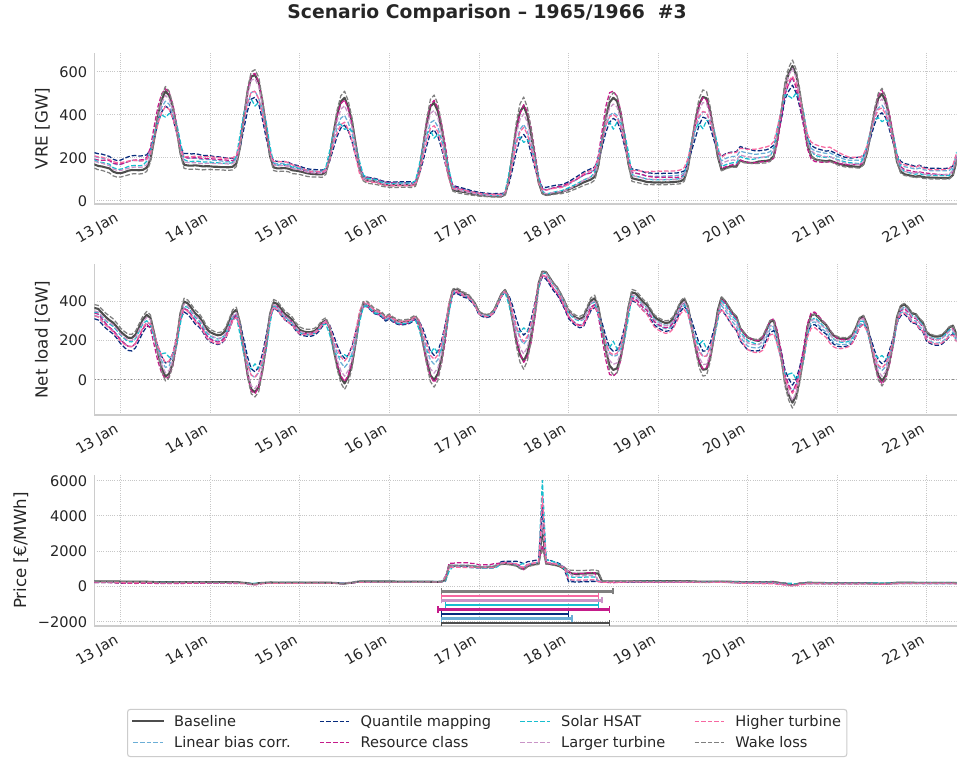}
  \caption{Stress event \#3 (weather year 1965/66): as
    Fig.~\ref{fig:stress_event_1}.}
  \label{fig:stress_event_3}
\end{figure}

\begin{figure}[htbp]
  \centering
  \includegraphics[width=0.89\linewidth]{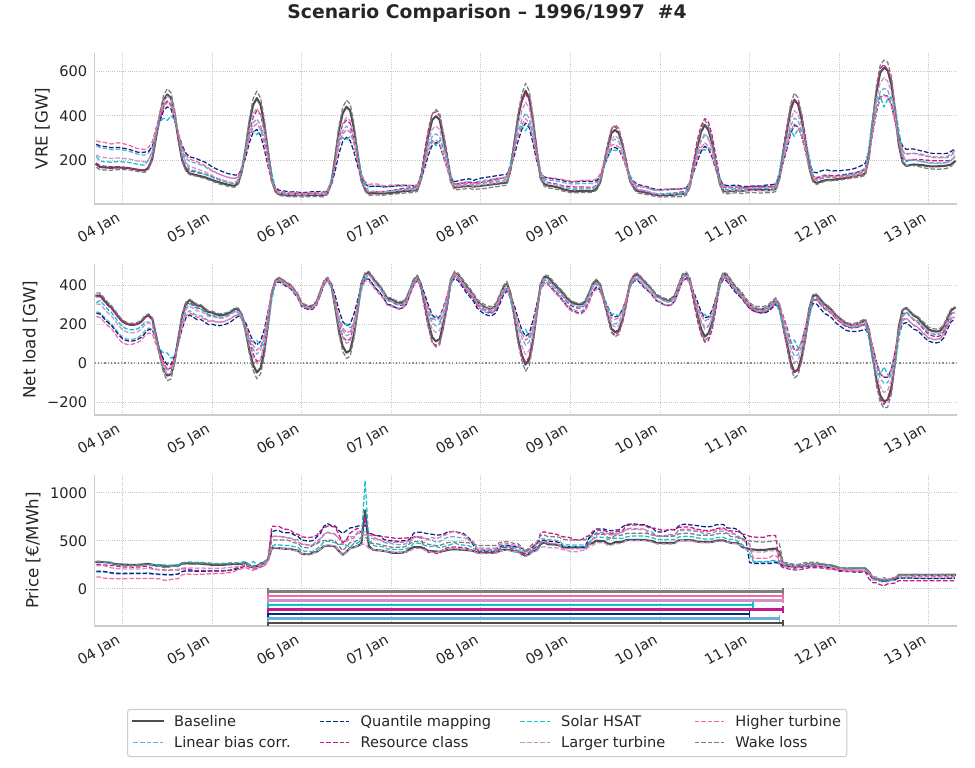}
  \caption{Stress event \#4 (weather year 1996/97): as
    Fig.~\ref{fig:stress_event_1}.}
  \label{fig:stress_event_4}
\end{figure}

\begin{figure}[htbp]
  \centering
  \includegraphics[width=0.89\linewidth]{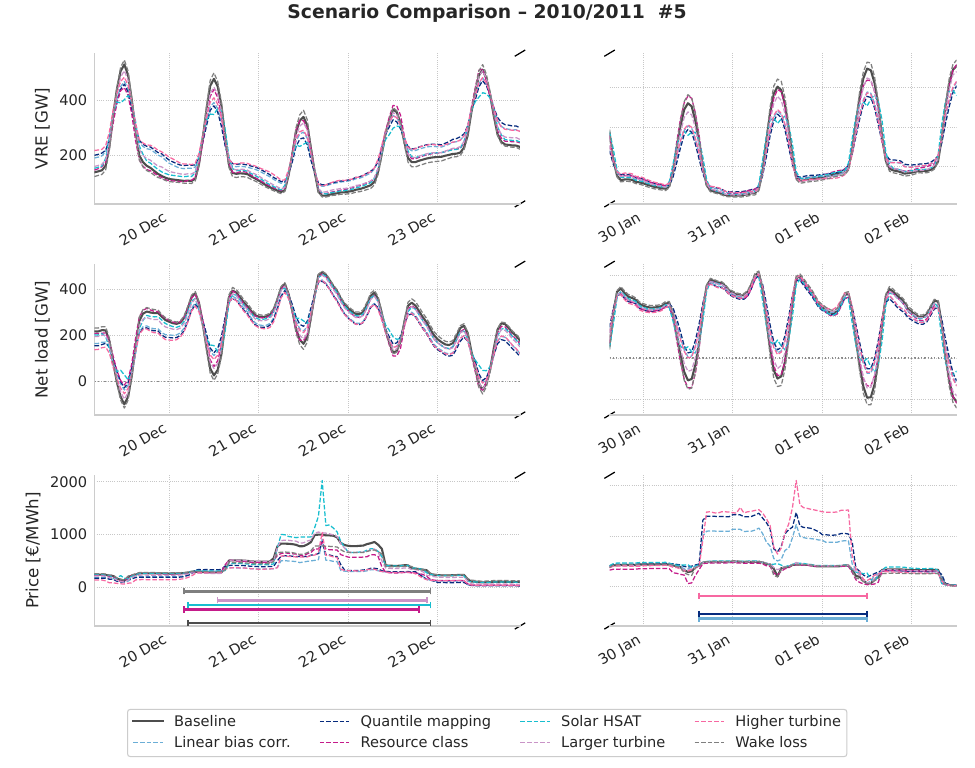}
  \caption{Stress event \#5 (weather year 2010/11): as
    Fig.~\ref{fig:stress_event_1}.}
  \label{fig:stress_event_5}
\end{figure}

Figures~\ref{fig:stress_event_1}--\ref{fig:stress_event_5}
show hourly VRE generation, net load, and electricity prices around each event;
horizontal bars indicate the severe period detected in each configuration.

\subsection{Battery discharge and H\texorpdfstring{\textsubscript{2}}{2}-turbine energy during stress events}
\label{sec:supp_battery_h2}
\begin{figure}[htbp]
  \centering
  \includegraphics[width=0.85\linewidth]{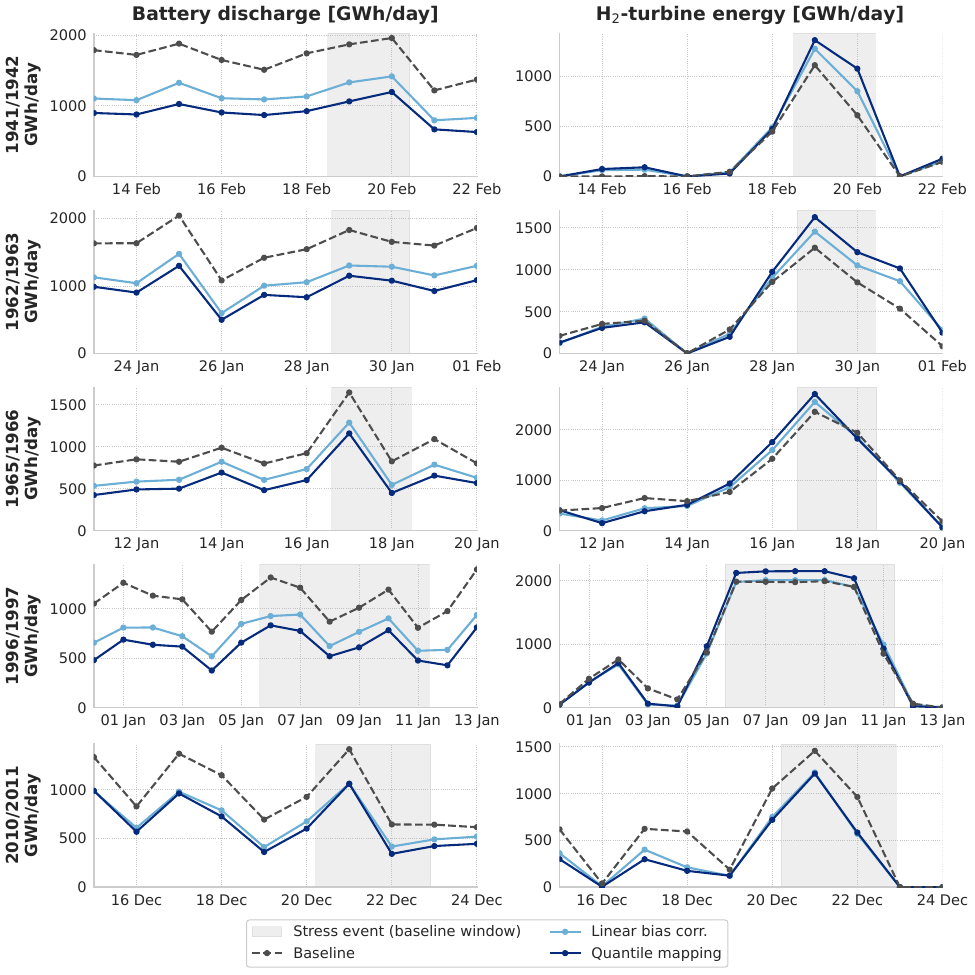}
  \caption{Daily battery discharge (left) and H\textsubscript{2}-turbine energy (right) before, during, and after each of the five system stress events for the baseline, linear correction, and quantile mapping. All configurations are shown over the same calendar window: the baseline's detected stress event (grey band) plus five days before and two days after. Battery discharge is lower under both bias corrections on almost every day, as their battery fleets are \qtyrange{30}{37}{\percent} smaller, whereas their H\textsubscript{2}-turbine energy is equal to or higher than the baseline's. Quantile mapping shows the lowest battery discharge and, except in 2010/11, uses the most H\textsubscript{2}-turbine energy. In 2010/11, however, both corrected systems use less H\textsubscript{2}-turbine energy than the baseline.}
  \label{fig:stress_battery_h2_daily}
\end{figure}

Figure~\ref{fig:stress_battery_h2_daily} complements the stress-event analysis in Sec.~\ref{sec:results_stress_events} by showing how the \qtyrange{30}{37}{\percent} smaller battery fleets under bias correction (Sec.~\ref{sec:results_system_configuration}) change the backup supply during the five events. Under both corrections, batteries discharge less, and H\textsubscript{2} turbines supply more energy than in the baseline, except in 2010/11.

